\documentclass[sigconf,screen]{acmart}
\usepackage{xspace,balance,tabularx,multirow}
\usepackage{flushend}
\usepackage{tikz}
\usepackage{pgfplots}
\pgfplotsset{compat=1.16}
\usetikzlibrary{patterns}
\usepackage{subfig}
\usepackage[ruled, vlined, linesnumbered]{algorithm2e}
\usepackage{xcolor}
\usepackage{colortbl}
\usepackage{bbold}
\SetKwComment{Comment}{$\triangleright$\ }{}
\usepackage{enumitem}
\usepackage{tablefootnote}
\usepackage{upgreek,textgreek}
\usepackage{pifont}
\usepackage[noabbrev]{cleveref}
\usepackage{titlecaps}
\usepackage{lipsum}

\pgfplotsset{every tick label/.append style={font=\tiny}}

\newlength{\starsize}
\newlength{\starspread}
\tikzset{starsize/.code={\setlength{\starsize}{#1}},
         starspread/.code={\setlength{\starspread}{#1}}}
\tikzset{starsize=1mm,
         starspread=3mm}
\pgfdeclarepatternformonly[\starspread,\starsize]
  {my fivepointed stars}
  {\pgfpointorigin}
  {\pgfqpoint{\starspread}{\starspread}}
  {\pgfqpoint{\starspread}{\starspread}}
  {
   \pgftransformshift{\pgfqpoint{\starsize}{\starsize}}
   \pgfpathmoveto{\pgfqpointpolar{18}{\starsize}}
   \pgfpathlineto{\pgfqpointpolar{162}{\starsize}}
   \pgfpathlineto{\pgfqpointpolar{306}{\starsize}}
   \pgfpathlineto{\pgfqpointpolar{90}{\starsize}}
   \pgfpathlineto{\pgfqpointpolar{234}{\starsize}}
   \pgfpathclose%
   \pgfusepath{fill}
  }

\newcommand{\argmax}[1]{\underset{#1}{\operatorname{arg}\,\operatorname{max}}\;}

\makeatletter
\newcommand*\bigcdot{\mathpalette\bigcdot@{.5}}
\newcommand*\bigcdot@[2]{\mathbin{\vcenter{\hbox{\scalebox{#2}{$\m@th#1\bullet$}}}}}
\makeatother

\newcommand{\stitle}[1]{\vspace*{0.5em}\noindent{\bf #1.\/}}

\newcommand{\V}{\mathcal{V}\xspace}
\newcommand{\G}{\mathcal{G}\xspace}
\newcommand{\N}{\mathcal{N}\xspace}
\newcommand{\EDG}{\mathcal{E}\xspace}
\newcommand{\C}{\mathcal{C}\xspace}

\newcommand{\WM}{\boldsymbol{W}\xspace}
\newcommand{\AM}{\boldsymbol{A}\xspace}
\newcommand{\DM}{\boldsymbol{D}\xspace}
\newcommand{\IM}{\boldsymbol{I}\xspace}
\newcommand{\SM}{\boldsymbol{S}\xspace}
\newcommand{\MM}{\boldsymbol{M}\xspace}
\newcommand{\NM}{\boldsymbol{N}\xspace}
\newcommand{\PM}{\boldsymbol{P}\xspace}
\newcommand{\CM}{\boldsymbol{C}\xspace}
\newcommand{\YM}{\boldsymbol{Y}\xspace}
\newcommand{\XM}{\boldsymbol{X}\xspace}
\newcommand{\LM}{\boldsymbol{L}\xspace}
\newcommand{\UM}{\boldsymbol{U}\xspace}
\newcommand{\VM}{\boldsymbol{V}\xspace}
\newcommand{\HM}{\boldsymbol{H}\xspace}
\newcommand{\ZM}{\boldsymbol{Z}\xspace}
\newcommand{\RM}{\boldsymbol{R}\xspace}
\newcommand{\FM}{\boldsymbol{F}\xspace}

\newcommand{\PiM}{\boldsymbol{\Pi}\xspace}

\newcommand{\NAM}{\boldsymbol{\hat{A}}\xspace}

\newcommand{\BM}{\boldsymbol{B}\xspace}
\newcommand{\TM}{\boldsymbol{T}\xspace}

\newcommand{\QM}{\boldsymbol{Q}\xspace}
\newcommand{\qvec}{\boldsymbol{q}\xspace}
\newcommand{\wvec}{\boldsymbol{\omega}\xspace}
\newcommand{\zvec}{\boldsymbol{z}\xspace}

\newcommand{\xvec}{\boldsymbol{x}\xspace}

\newcommand{\algo}{\textsf{S\textsuperscript{2}CAG}\xspace}
\newcommand{\algoplus}{\textsf{M-S\textsuperscript{2}CAG}\xspace}
\newcommand{\itrgf}{\textsf{PowerMethod}\xspace}

\newenvironment{customlegend}[1][]{%
    \begingroup
    \csname pgfplots@init@cleared@structures\endcsname
    \pgfplotsset{#1}%
}{%
    \csname pgfplots@createlegend\endcsname
    \endgroup
}%

\def\addlegendimage{\csname pgfplots@addlegendimage\endcsname}

\makeatletter
\newcommand\footnoteref[1]{\protected@xdef\@thefnmark{\ref{#1}}\@footnotemark}
\makeatother

\let\oldnl\nl
\newcommand{\nonl}{\renewcommand{\nl}{\let\nl\oldnl}}

\SetKwComment{Comment}{/* }{ */}

\makeatletter 
\g@addto@macro{\@algocf@init}{\SetKwInOut{Parameter}{Parameters}} 
\makeatother

\definecolor{myred}{HTML}{fd7f6f}
\definecolor{myred_new}{HTML}{D8D8D8}
\definecolor{myred_new2}{HTML}{D7191C}
\definecolor{myblue}{HTML}{7eb0d5}
\definecolor{mygreen}{HTML}{b2e061}
\definecolor{mypurple}{HTML}{bd7ebe}
\definecolor{myorange}{HTML}{ffb55a}
\definecolor{myyellow}{HTML}{ffee65}
\definecolor{mypurple2}{HTML}{beb9db}
\definecolor{mypink}{HTML}{fdcce5}
\definecolor{mycyan}{HTML}{8bd3c7}

\definecolor{myblue2}{HTML}{115f9a}
\definecolor{myred2}{HTML}{c23728}

\newcommand{\eat}[1]{}

\AtBeginDocument{%
  \providecommand\BibTeX{{%
    \normalfont B\kern-0.5em{\scshape i\kern-0.25em b}\kern-0.8em\TeX}}}

\copyrightyear{2025}
\acmYear{2025}
\setcopyright{rightsretained}
\acmConference[KDD '25]{Proceedings of the 31th ACM SIGKDD Conference on Knowledge Discovery and Data Mining}{August 3--7, 2025}{Toronto, Canada}
\acmBooktitle{Proceedings of the 31th ACM SIGKDD Conference on Knowledge Discovery and Data Mining (KDD '25), August 3--7, 2025, Toronto, Canada}
\acmDOI{XXXXXXX.XXXXXXX}
\acmISBN{978-1-4503-XXXX-X/18/06}

\begin{document}


\title{Spectral Subspace Clustering for Attributed Graphs}

\author{Xiaoyang Lin}
\affiliation{%
  \institution{Hong Kong Baptist University}
  \country{}
}
\email{csxylin@hkbu.edu.hk}

\author{Renchi Yang}
\affiliation{%
  \institution{Hong Kong Baptist University}
  \country{}
}
\email{renchi@hkbu.edu.hk}

\author{Haoran Zheng}
\affiliation{%
  \institution{Hong Kong Baptist University}
  \country{}
}
\email{cshrzheng@hkbu.edu.hk}

\author{Xiangyu Ke}
\affiliation{%
  \institution{Zhejiang University}
  \country{}
}
\email{xiangyu.ke@zju.edu.cn}

\renewcommand{\shortauthors}{Lin et al.}

\begin{abstract}
Subspace clustering seeks to identify subspaces that segment a set of $n$ data points into $k$ ($k\ll n$) groups, which has emerged as
a powerful tool for analyzing data from various domains, especially images and videos.
Recently, several studies have demonstrated the great potential of subspace clustering models for partitioning vertices in attributed graphs, referred to as SCAG.
However, these works either demand significant computational overhead for constructing the $n\times n$ self-expressive matrix, or fail to incorporate graph topology and attribute data into the subspace clustering framework effectively, and thus, compromise result quality.

Motivated by this, this paper presents 
two effective and efficient algorithms, \algo and \algoplus, for SCAG computation. Particularly, \algo obtains superb performance through three major contributions. 
First, we formulate a new objective function for SCAG with a refined representation model for vertices and two non-trivial constraints.
On top of that, an efficient linear-time optimization solver is developed based on our theoretically grounded problem transformation and well-thought-out adaptive strategy. We then conduct an in-depth analysis to disclose the theoretical connection of \algo to conductance minimization, which further inspires the design of \algoplus that maximizes the modularity.
Our extensive experiments, comparing \algo and \algoplus against 17 competitors over 8 benchmark datasets, exhibit that our solutions outperform all baselines in terms of clustering quality measured against the ground truth while delivering high efficiency.
\end{abstract}




\maketitle

\section{Introduction}
{\em Attributed graphs} are an omnipresent data structure used to model the interplay between entities characterized by rich attributes, such as social networks, citation graphs, World Wide Web, and transportation grids.
Clustering such graphs, i.e., partition vertices therein into disjoint groups, has gained massive attention over the past decade~\cite{bothorel2015clustering,chunaev2020community,liu2022survey}, due to its encouraging performance by 
leveraging the complementary nature of graph topology and attributes, as well as its extensive use in community detection~\cite{tang2008arnetminer,liu2024dag}, Bioinformatics~\cite{cheng2022scgac,buterez2022cellvgae}, anomaly identification~\cite{luo2022comga}, online advertising and recommendation~\cite{liji2018improved,eissa2018towards}, etc.

This problem remains tenaciously challenging as it requires joint modeling of graph structures and nodal attributes that present heterogeneity, complex semantics, and noises.
State-of-the-art solutions~\cite{lai2023re,ding2024towards} are built upon deep learning techniques, especially {\em graph convolutional neural networks}~\cite{kipf2022semi},
which first aggregates attributes of neighbors in the graph before mapping attribute vectors to low-dimensional feature representations of vertices for clustering.
Notwithstanding promising results reported, these works rely on a strong assumption that neighboring vertices should have high attribute homogeneity, rendering them vulnerable to attributed networks contaminated with irrelevant and noisy data~\cite{zhu2020beyond}.
In addition, this category of methods suffers from poor scalability and demands tremendous computational resources for model training, especially on large graphs.

{\em Subspace clustering} (SC)~\cite{vidal2011subspace} is a fundamental technique in data mining used for analyzing high-dimensional data including images, text, 
gene expression data, 
and so on~\cite{elhamifar2013sparse}. 
Intrinsically, it 
simultaneously clusters the data into multiple subspaces and identifies a low-dimensional subspace fitting each group of points, thereby eradicating the adverse impacts of irrelevant and noisy dimensions (i.e., features)~\cite{parsons2004subspace}.
More precisely, the linchpin of SC is to construct a {\em self-expressive matrix} (SEM) that models the affinity of all data points such that each data point can be written as a linear combination of others.
Subsequent research~\cite{gao2015multi,zhang2018generalized} further bolsters the SC performance by working in tandem with {\em multi-view data} where the data set is represented by multiple distinct feature sets.

Inspired by its success, in recent years, several attempts~\cite{fettal2023scalable,ma2020towards,li2024attributed} have been made towards extending the principle of SC to attributed graphs for enhanced clustering quality, given that an attributed graph embodies two sets of features (i.e., structures and attributes).
For simplicity, we refer to this line of research as SCAG (short for subspace clustering for attributed graphs) henceforth.
Among these works, \textsf{SAGSC}~\cite{fettal2023scalable} heavily relies on feature augmentation via polynomial combinations of the extracted features. Such hand-crafted features decrease the model interpretability and its generalization to diverse datasets and engender suboptimal result quality.
\citet{ma2020towards} and Li et al.~\cite{li2024attributed} directly construct the SEM, resulting in space and computation costs quadratic in the number of vertices. As an aftermath, they struggle to cope with real attributed graphs, which often encompass numerous vertices, edges, and a sheer volume of attribute data. Furthermore, despite their empirical effectiveness on attributed graphs, there is little theoretical interpretation and analysis of these SCAG approaches.

To overcome the deficiencies of existing solutions, this paper presents \algo and \algoplus which offer superb clustering performance while being efficient, through a series of novel algorithmic designs and rigorous theoretical analyses.
First and foremost, \algo{} formulates the optimization objective of SCAG based on our {\em normalized smoothed representations} (NSR) for vertices with a {\em low-rank constraint}~\cite{Liu2010RobustSS} and an orthogonality regularization. Particularly, NSR enables stronger representation capacity by augmenting {\em graph Laplacian smoothing}-based vertex representations~\cite{dong2016learning} with normalized topology, attribute matrices, and weights. The two additional constraints improve the resulting SEM's robustness to noises and outliers, and meanwhile, facilitate the design of our efficient optimization solvers.
Taking inspiration from the {\em orthogonal Procruste problem}~\cite{Schnemann1966AGS}, \algo transforms the problem into a simple truncated {\em singular value decomposition} (SVD)~\cite{Halko2009FindingSW} without materializing the SEM, whereby we devise adaptive solvers that merely involve efficient operations on tall-and-skinny matrices.
Moreover, we conduct detailed theoretical analyses digesting \algo's relation to graph clustering that optimizes {\em conductance}~\cite{lovasz1993random}.
This finding further guides us to upgrade \algo to \algoplus which essentially maximizes {\em modularity}~\cite{Newman2006ModularityAC} on affinity graphs for clustering.

Our empirical studies, which evaluate \algo and \algoplus against 17 baselines over 8 real-life attributed graphs with ground-truth clusters, showcase the consistent superiority of our solutions in terms of result quality and efficiency. On the {\em ArXiv} dataset with 169K vertices and 1.3M edges, \algoplus advances the state of the art by a significant improvement of $2.2\%$ in clustering accuracy without degrading the practical efficiency.

\section{Related Work}

\subsection{Subspace Clustering}
Foundational subspace clustering algorithms like Sparse Subspace Clustering ({SSC})~\cite{Elhamifar2012SparseSC}, Low-Rank Representation ({LRR})~\cite{Liu2010RobustSS}, and Multiple Subspace Representation ({MSR})~\cite{Luo2011MultiSubspaceRA} focus on learning an affinity matrix, which is then employed in spectral clustering to discern underlying data structures.

However, the traditional methods often grapple with high computational complexity. In response, a suite of low-rank sparse methods has been proposed~\cite{Lu2012Lsr,Chen2014SubspaceCU,You2015ScalableSS,Fan2021kfsc}, which simultaneously reduce computational load and improve algorithmic efficiency and scalability. Further enhancements in subspace clustering algorithms have been achieved through methods that emphasize local structure-preserving and kernel techniques~\cite{Lu2013GraphRegularizedLR,Kang2021StructuredGL,Patel2014KernelSS,Liu2021AdaptiveLK}. For an in-depth exploration, we kindly refer to surveys~\cite{vidal2011subspace,kelkar2019subspace,qu2023survey}.
Furthermore, a number of studies have successfully incorporated deep learning into the traditional subspace clustering paradigm ~\cite{Ji2017DeepSC,Zhou2018DeepAS,Zhou2019LatentDP,peng2020deep,Cai2022EfficientDE}. These approaches leverage end-to-end training to learn proximities in latent subspaces, ultimately producing a representative coefficient matrix that is critical for subsequent clustering processes.

\subsection{Attributed Graph Clustering}
Extensive studies have identified various methods for addressing attribute graph clustering (\textsf{AGC}) challenges. These include approaches grounded in edge-weight-based~\cite{Combe2012CombiningRA,Meng2018CoupledNS,Neville2003ClusteringRD,Ruan2012EfficientCD,Steinhaeuser2008CommunityDI}, distance-based~\cite{Nawaz2012CollaborativeSM,Falih2017ANCAA,Zhu2021SimpleSG} and probabilistic-model-based methods~\cite{Yang2014CombiningLA,Zanghi2009ClusteringBO,Xu2012AMA,Nowicki2001EstimationAP}. For detailed insights, readers could refer to related reviews ~\cite{bothorel2015clustering,yang2021effective,chunaev2020community,li2024versatile}.

Recently, common practice involves integrating structural connectivity into vertex attributes to derive vertex embeddings~\cite{Akbas2017AttributedGC,Li2018CommunityDI,Combe2015ILouvainAA,Zhu2021SimpleSG,yang2023pane}, which are then used to perform clustering via established techniques such as \textsf{KMeans}. Lai et al.\cite{lai2023re} conducted a comprehensive assessment of all existing deep attribute graph clustering (\textsf{DAGC}) methods~\cite{Tu2020DeepFC,Hao2022DeepGC,Huo2021CaEGCNCF,Cui2020AdaptiveGE, sdcn2020, DCRN} for \textsf{AGC}, which capture both the topological and attribute information of graphs, integrating this fused information to facilitate the learning of vertex embeddings. To augment the proficiency of vertex representation learning, certain models embedded in the \textsf{DAGC} framework have embraced graph attention mechanisms \cite{Xia2023RobustCM,Zhao2022HierarchicalAN,Zhou2022CommunityDB, wang2019attributed}, coupled with advanced graph contrastive learning methodologies \cite{Yang2023ClusterguidedCG,Hassani2020ContrastiveMR,Zhao2021GraphDC}. Beyond the enhancement of vertex representations to achieve superior clustering performance, certain models \cite{Mavromatis2021GraphIM,fettal2022efficient,liu2023dink} incorporate the outcomes of vertex clustering into deep learning frameworks to optimize the cluster distribution.

Owing to the high efficiency demonstrated by subspace clustering algorithms, a growing body of research~\cite{gunnemann2013spectral,ma2020towards,li2024attributed,wei2023adaptive} has been applying these algorithms to \textsf{AGC}. These methodologies perform SC algorithms on graphs that integrate both topological and attribute information. Among these, \textsf{SAGSC} \cite{fettal2023scalable} stands out as a state-of-the-art algorithm, characterized by its scalability and efficiency in SC. Nevertheless, the previously mentioned methodologies are constrained by their incapacity to thoroughly exploit the graph's topological framework and nodal attribute data, concomitant with an absence of low-complexity SC underpinned by stringent mathematical principles.

\section{Problem Formulation}

\subsection{Notations and Terminology}


Throughout this paper, sets are symbolized by calligraphic letters, e.g., $\V$. Matrices (resp. vectors) are denoted as bold uppercase (resp. lowercase) letters, e.g., $\XM$ (resp. $\xvec$). The transpose and inverse of matrix $\XM$ are denoted by $\XM^{\top}$ and $\XM^{-1}$, respectively. The $i$-th row (resp. column) of $\XM$ is represented by $\XM_i$ (resp. $\XM_{\cdot,i}$). Accordingly, $\XM_{i,j}$ stands for the $(i,j)$-th entry of $\XM$. $\|\XM\|_F$ stands for the Frobenius norm of matrix $\XM$. We use $\IM$ to denote the identity matrix and its size is obvious from context. 
We refer to the left (resp. right) singular vectors of $\XM$ that correspond to its $k$-largest singular values as {\em top-$k$ left (resp. right) singular vectors}. The $k$ eigenvectors of $\XM$ corresponding to its $k$ largest eigenvalue in absolute value are referred to as the {\em $k$-largest eigenvectors of $\XM$}.

An {\em attributed graph} is defined as $\G=(\V,\EDG,\XM)$, composed of a set $\V=\{v_1,v_2,\ldots,v_n\}$ of $n$ vertices (a.k.a. nodes), a set $\EDG\subseteq \V\times \V$ of $m$ edges, and an $n\times d$ attribute matrix $\XM$. Each vertex $v_i\in \V$ is characterized by a length-$d$ attribute vector $\XM_i$ and each edge $(v_i,v_j)\in \EDG$ connects two vertices $v_i,v_j\in \V$. For each vertex $v_i$, we denote by $\N_{v_i}=\{v_j\in \V| (v_i,v_j)\in \EDG\}$ the set of direct neighbors of $v_i$, and by $d(v_i):=|\N_{v_i}|$ the degree of $v_i$. We use $\AM\in \{0,1\}^{n\times n}$ to represent the adjacency matrix (with self-loops) of $\G$, where $\AM_{i,j}=1$ if $(v_i,v_j)\in \EDG$ or $v_i=v_j$, and 0 otherwise. In this paper, we consider undirected graphs, and hence, $\AM$ is symmetric. The diagonal degree matrix of $\G$ is symbolized by $\DM$, wherein the diagonal entry $\DM_{i,i}=d(v_i)+1$. The normalized adjacency matrix $\NAM$ and transition matrix $\PM$ of $\G$ are defined as $\DM^{-\frac{1}{2}}\AM\DM^{-\frac{1}{2}}$ and $\DM^{-1}\AM$, respectively. In particular, $\PM^t_{i,j}$ denotes the probability that a $t$-hop simple random walk from vertex $v_i$ would stop at vertex $v_j$. Accordingly, the {\em personalized PageRank} (PPR)~\cite{jeh2003scaling} of vertex $v_j$ w.r.t. vertex $v_i$ over $\G$ is defined as $\textstyle \pi_{v_i}(v_j)=\sum_{t=0}^{\infty}{(1-\alpha)\alpha^t\PM^t_{i,j}}$, where $\alpha$ stands for the decay factor.

\eat{
Let $\G=(\V,\EDG,\XM)$ be an {\em attributed graph} wherein $\V$ represents a set of $n$ nodes and $\EDG$ contains $m$ edges between nodes. For each edge $e_{i,j}\in \EDG$, it denotes that $v_i$ and $v_j$ are neighbors to each other. $\AM \in \mathbb{R}^{n\times n} $ is a symmetric adjacency matrix, the elements $a_ij$ in this matrix are the edge weights between $v_i$ and $v_j$, $\XM\in \mathbb{R}^{n\times d}$ is a node-level feature matrix, while $d$ is the feature dimension of each node. Specifically, trace denotes the trace of matrix, and We use k to denote the number of clusters that need to be partitioned for each graph. If  $\UM \in \mathbb{R}^{n\times n} $, $m_i$ is the i-th row vector of $\UM$ and ${m}_{j}^{'}$ means the j-th columns vector of $\UM$, then $\UM_{i,j}$ can be interpreted as taking the first i rows and the first j columns of the matrix $\UM$ as a new matrix. I represent the identity matrix. 
and we use $\N(v_i)$ to denote the set of neighbors of $v_i$, where the degree is $d(v_i)=|\N(v_i)|$. 
}

\subsection{Graph Laplacian Smoothing}\label{se:gs}
Given a graph $\G$ and a signal vector $\xvec\in \mathbb{R}^n$, the goal of {\em graph Laplacian smoothing}~\cite{dong2016learning} is to find a smoothed version of $\xvec$, i.e., $\overline{\xvec}\in\mathbb{R}^n$, such that the following objective is optimized:
\begin{equation}\label{eq:GLS-vec}
\min_{\overline{\xvec}}{\|\overline{\xvec} - \xvec\|^2_2} + \alpha \cdot \overline{\xvec}^\top \LM \overline{\xvec},
\end{equation}
where $\LM=\IM-\NAM$ denotes the normalized Laplacian matrix of $\G$ and $\alpha\in (0,1)$ is the parameter balancing two terms.
If we consider $d$ different signal vectors, e.g., the attribute matrix $\XM\in \mathbb{R}^{n\times d}$, the overall optimization objective is therefore
\begin{equation}\label{eq:GLS-mat}
\min_{\overline{\XM}}{(1-\alpha)\cdot\|\overline{\XM} - \XM\|^2_F} + \alpha \cdot trace(\overline{\XM}^\top \LM \overline{\XM}),
\end{equation}
The first term in Eq.~\eqref{eq:GLS-mat} seeks to reduce the discrepancy between the input matrix $\XM$ and its smoothed version $\overline{\XM}$. 
By~\cite{von2007tutorial}, $trace(\overline{\XM}^\top \LM \overline{\XM})$ can be rewritten as $\small \textstyle\sum_{(v_i,v_j)\in \EDG}{\left\|\frac{\overline{\XM}_{i}}{\sqrt{d(v_i)}}-\frac{\overline{\XM}_{j}}{\sqrt{d(v_j)}}\right\|^2_2}$, 
meaning that the second term enforces the smoothed attribute vectors $\overline{\XM}_i$, $\overline{\XM}_j$ of any adjacent vertices $(v_i,v_j)\in \EDG$ to be similar. 

\begin{lemma}[Neumann Series~\cite{horn2012matrix}]\label{lem:NM}
If the eigenvalue $\lambda_i$ of $\MM \in \mathbb{R}^{n\times n}$ satisfies $|\lambda_i|<1\ \forall{1\le i\le n}$, then $(\IM-\MM)^{-1}=\sum_{t=0}^{\infty}{\MM^{t}}$.
\end{lemma}

As demystified in recent studies~\cite{Ma2020AUV,Zhu2021InterpretingAU}, after removing non-linear operations, graph convolutional layers in popular {\em graph neural network} (GNN) models, e.g., APPNP~\cite{gasteiger2018predict}, GCNII~\cite{chen2020simple}, 
essentially optimize the objective in Eq.~\eqref{eq:GLS-mat} and the closed-form solution (i.e., final vertex representations) can be represented as
\begin{small}
\begin{equation}\label{eq:smooth-x}
\textstyle \overline{\XM} = (1-\alpha)\cdot \left((1-\alpha)\cdot\IM+\alpha\cdot \LM\right)^{-1}\XM = \sum_{t=0}^{\infty}{(1-\alpha)\alpha^t\NAM^t}\XM
\end{equation}
\end{small}
by taking the derivative of Eq.~\eqref{eq:GLS-mat} with respect to $\overline{\XM}$ to zero and applying Lemma~\ref{lem:NM} with $\MM=\alpha\cdot \NAM$ ($\alpha\in (0,1)$). 
Since $\NAM^t = \DM^{\frac{1}{2}}\PM^{t}\DM^{-\frac{1}{2}}$, we can rewrite $\sum_{t=0}^{\infty}{\alpha^t\NAM^t}$ as $\DM^{\frac{1}{2}}\cdot\left(\sum_{t=0}^{\infty}{\alpha^t\PM^t}\right)\cdot\DM^{-\frac{1}{2}}$.
As such, the smoothed representation $\overline{\XM}_i$ of each vertex $v_i\in \V$ obtained in Eq.~\eqref{eq:smooth-x} is a summation of the attribute vectors of all vertices weighted by their degree-reweighted PPR w.r.t. $v_i$, i.e., 
\begin{small}
\begin{equation}\label{eq:x-ppr}
\textstyle \overline{\XM}_i = \sum_{v_j\in \V}{\sqrt{\frac{d(v_i)}{d(v_j)}}\cdot \pi_{v_i}(v_j) \cdot \XM_j}.
\end{equation}
\end{small}
In practice, $\overline{\XM}$ in Eq.~\eqref{eq:smooth-x} is usually approximated via a $T$-order truncated version $\textstyle \sum_{t=0}^{T}{(1-\alpha)\alpha^t\NAM^t}\XM$, where $T$ is typically dozens.

\subsection{Subspace Clustering}

Let $\FM\in \mathbb{R}^{n\times d}$ be a data matrix for $n$ distinct data samples where each data sample $i\in \{1,2,\ldots,n\}$ is represented by a $d$-dimensional feature vector $\FM_i$. {\em Subspace clustering} aims to group data samples into $k$ disjoint clusters $\{\C_1,\C_2,\ldots,\C_k\}$, which is based on the assumption that data samples lie in a union of subspaces~\cite{vidal2011subspace}. {\em Self-Expression Model}~\cite{Lu2012Lsr} is the most widely adopted objective formulation for subspace clustering. In this model, each data sample is assumed to be expressed as a linear combination of other data samples in the same subspace:
\begin{equation}\label{eq:subspace-obj}
\min_{\SM\in \mathbb{R}^{n\times n}}{\|\FM-\SM\FM\|_F^2}+\Omega(\SM),
\end{equation}
where $\SM\in \mathbb{R}^{n\times n}$ is known as the {\em self-expressive matrix} (SEM) (a.k.a. {\em coefficient matrix}). The first term is to reconstruct $\FM$ via $\SM$ and $\FM$, while the regularization term $\Omega(\SM)$ is introduced to 
impose constraints rendering $\SM$ meet certain structures or averting trivial solutions, e.g., $\IM$.
Popular constraints include {\em sparsity constraint}~\cite{wang2013provable} and {\em low-rank representation} (LRR)~\cite{Liu2010RobustSS}. The former minimizes the vector $L_1$ norm of $\SM$ to induce sparsity, whereas LRR minimizes 
the rank of $\SM$ such that $\SM$ captures the global correlation of the data samples~\cite{sui2019sparse}.
In simpler terms, with the low-rank constraint, correlations between data samples are strengthened within clusters but weakened across clusters. 
Besides, by virtue of the low-rank setting, we can extract dominant patterns/features in the data while filtering out minor deviations, and hence, improve the robustness to noise and outliers.
The resulting SEM $\SM$ is then used to form an affinity matrix $\frac{\SM+\SM^{\top}}{2}$ that quantifies the affinity of every two data samples.
Based thereon, 
{\em spectral clustering}~\cite{von2007tutorial} can be applied to the affinity matrix for clustering.

\subsection{Subspace Clustering for Attributed Graphs}\label{sec:SCAG}
To extend subspace clustering to attributed graphs, a simple and straightforward idea is to employ the vertex representations $\overline{\XM}$ obtained via GLS in Eq.~\eqref{eq:smooth-x} as the data feature matrix $\FM$.

\stitle{Normalized Smoothed Representations} 
We argue that the direct adoption of $\overline{\XM}$ for subspace clustering is problematic. First, $\alpha$ in $\overline{\XM}$ (Eq.~\eqref{eq:smooth-x}) is restricted within the range $(0,1)$ to avoid negative or zero values due to the existence of $1-\alpha$. Although $1-\alpha$ can be removed, we still cannot assign large values to $\alpha$ as it leads to weighty coefficients $\alpha^t$ that might overwhelm the entries in $\NAM^t$ and other terms. 
Thereby, $\alpha^t$ is constrained to monotonically decrease as $t$ increases, limiting the capacity of $\overline{\XM}$ in capturing the topological semantics in various graphs. Moreover, each entry $\NAM_{i,j}\ \forall{v_i,v_j\in \V}$ merely considers the degrees of endpoints $v_i$ and $v_j$ and overlooks the structures of other adjacent vertices of $v_i$ or $v_j$, which is apt to cause biased attribute aggregation in Eq.~\eqref{eq:x-ppr}. Similar issues arise on $\XM$, where attribute vectors of vertices fall on different scales. In response, 
we propose to calculate the {\em normalized smoothed representations} (NSR) of vertices in $\G$ as $\FM$ via
\begin{small}
\begin{equation}\label{eq:NSR}
\textstyle \ZM = \sum_{t=0}^{T}{\frac{\alpha^t}{\sum_{\ell=0}^{T}{\alpha^\ell}}\hat{\PM}^{t}}\hat{\XM} = \sum_{t=0}^{T}{\frac{(1-\alpha)\alpha^{t}}{1-\alpha^{\TM+1}}\hat{\PM}^{t}}\hat{\XM},
\end{equation}
\end{small}
where an $L_1$ normalization is applied to all weights $1,\alpha,\ldots,\alpha^T$ such that $\alpha$ is allowed to exceed $1$. As for $\NAM$ and $\XM$ in Eq.~\eqref{eq:smooth-x}, we substitute them by their normalized versions $\hat{\PM}$ and $\hat{\XM}$ defined by
\begin{equation*}
\textstyle \hat{\PM}_{i,j} = \frac{\NAM_{i,j}}{\sum_{v_\ell\in \V}\NAM_{i,\ell}}\ \forall{v_i,v_j\in \V}\ \text{and}\ \hat{\XM}_i = \frac{\XM_i}{\sqrt{\sum_{v_j\in \V}\XM_i\cdot {\XM_j}^{\top}}}\ \forall{v_i\in \V}.
\end{equation*}

\stitle{Objective Function} Based on NSR $\ZM$ and Eq.~\eqref{eq:subspace-obj}, we impose the {\em low-rank constraint} (i.e., LRR) and {\em soft orthogonality regularization} to SEM $\SM$ and formulate the main objective function for SCAG as
\begin{equation}\label{eq:obj}
\min_{\SM\in \mathbb{R}^{n\times n}}{\|\ZM-\SM\ZM\|_F^2} + \|\SM\|_{*} + \|\SM^{\top}\SM-\IM\|_F^2,
\end{equation}
where the nuclear norm~\cite{fazel2002matrix} $\|\SM\|_{*}$ is an approximation of $rank(\SM)$ and the regularizer $\|\SM^{\top}\SM-\IM\|_F^2$ is introduced to 
prevent {\em overcorrelation} between vertices.

The above objective poses two formidable technical challenges. First and foremost, the resulting SEM $\SM$ is a dense matrix whose materialization consumes $O(n^2)$ space storage, which is prohibitive for large graphs. Using such a dense matrix for the rear-mounted spectral clustering further yields an exorbitant expense of $O(n^2k)$ time. On top of that, the complex constraints and objectives in Eq.~\eqref{eq:obj} lead to numerous iterations till convergence towards its optimization, each of which takes a quadratic time of $O(n^2)$ due to the high density of $\SM$.

\section{\algo Approach}\label{sec:algo}
To cope with the above-said issues, this section presents a practical algorithm \algo for SCAG that runs in time linear to the size of $\G$. Section~\ref{sec:problem-trans} transforms our optimization objective in Eq.~\eqref{eq:obj} into a simple truncated SVD of $\ZM$ based on a rigorous theoretical analysis. Section~\ref{sec:adap-y} delineates how \algo implements the SVD and clustering in an adaptive fashion for higher efficiency. In Section~\ref{sec:analysis}, we establish the theoretical connections between \algo and minimizing the {\em conductance}~\cite{lovasz1993random} of clusters.

\subsection{High-level Idea}\label{sec:problem-trans}
Note that the ultimate goal of SCAG is to partition $\V$ into $k$ disjoint clusters rather than computing the intermediate $\SM$. In light of this, we capitalize on the following theoretical analyses to bypass the explicit construction of $\SM$ and streamline the SCAG computation.

\stitle{Orthogonal Procrustes Transformation}
Instead of optimizing Eq.~\eqref{eq:obj} directly, we resort to an approximate solution by relaxing its constraints as follows. First of all, recall that minimizing the nuclear norm $\|\SM\|_{*}$ is to reduce the rank of $\SM$ as much as possible. However, the spectral clustering stage in SCAG requires computing the $k$ eigenvectors that correspond to the largest {\em non-zero} eigenvalues of $\SM$ or its Laplacian for clustering. Intuitively, an SEM $\SM$ with a rank $r<k$ has solely $r$ non-zero eigenvalues, which is incompetent for producing $k$ meaningful clusters.
Simply, we enforce the rank of $\SM$ to be $k$ as a trade-off.
With Lemma~\ref{lem:op}\footnote{All proofs can be found in Appendix~\ref{sec:proof}.}, our optimization objective in Eq.~\eqref{eq:obj} is transformed into an {\em orthogonal Procrustes problem}~\cite{Schnemann1966AGS} with a $k$-rank constraint over $\SM$ by letting $\MM=\NM=\ZM$ and $\boldsymbol{\Omega}=\SM$.
\begin{lemma}\label{lem:op}
Given matrices $\MM$ and $\NM$, the orthogonal Procrustes problem with a $k$-rank constraint aims to find matrix $\boldsymbol{\Omega}$ such that 
\begin{equation*}
\min_{\boldsymbol{\Omega}}{\|\boldsymbol{\Omega}\MM-\NM\|^2_F + \|\boldsymbol{\Omega}^{\top}\boldsymbol{\Omega}-\IM\|_F^2}\ \text{s.t.}\ rank(\boldsymbol{\Omega})=k.
\end{equation*}
The optimal value of $\boldsymbol{\Omega}$ is $\UM^{(k)}{\VM^{(k)}}^{\top}$, where the columns in $\UM^{(k)}$ and $\VM^{(k)}$ are the top-$k$ left and right singular vectors of $\NM\MM^{\top}$, respectively.
\end{lemma}
Consequently, an approximate minimizer $\SM$ to Eq.~\eqref{eq:obj} is $\UM^{(k)}{\VM^{(k)}}^{\top}$, where the columns of $\UM^{(k)}$ and $\VM^{(k)}$ are the top-$k$ left and right singular vectors of $\ZM\ZM^{\top}$, respectively.

\begin{lemma}\label{lem:obj-rank}
$\UM^{(k)}=\VM^{(k)}$ and their columns correspond to the top-$k$ left singular vectors of $\ZM$. 
\end{lemma}
Further, our careful analysis in Lemma~\ref{lem:obj-rank} reveals that SEM $\SM=\UM^{(k)}{\UM^{(k)}}^{\top}$, where $\UM^{(k)}$ denotes the top-$k$ left singular vectors of $\ZM$.
In turn, the problem in Eq.~\eqref{eq:obj} is reformulated as a simple $k$-truncated SVD of $\ZM$.


\stitle{Decomposed Spectral Clustering}
Recall that the second step of SCAG is to apply the spectral clustering to the affinity matrix $\frac{\SM+\SM^{\top}}{2}$, which first finds the $k$-largest eigenvectors $\YM\in \mathbb{R}^{n\times k}$ of affinity matrix $\frac{\SM+\SM^{\top}}{2}$, followed by a \textsf{$k$-Means} or rounding algorithms~\cite{shi2003multiclass,yang2024efficient,li2023efficient} over $\YM$ to recast it into a vertex-cluster assignment (VCA) matrix $\CM\in \mathbb{R}^{n\times k}$ (corresponding to the clusters $\{\C_1,\C_2,\ldots,\C_k\}$):
\begin{small}
\begin{equation}\label{eq:VCA}
\textstyle \CM_{i,j} =
\begin{cases}
\frac{1}{\sqrt{|\C_j|}} & \text{if $v_i\in \C_j$,}\\
0 & \text{Otherwise.}
\end{cases}
\end{equation}
\end{small}

Given that $\SM=\UM^{(k)}{\UM^{(k)}}^{\top}$ is a symmetric matrix, the affinity matrix $\frac{\SM + \SM^\top}{2}$ can be simplified as $\SM=\UM^{(k)}{\UM^{(k)}}^{\top}$ and the foregoing task turns to calculate the $k$-largest eigenvectors $\YM$ of $\UM^{(k)}{\UM^{(k)}}^{\top}$, which is exactly $\UM^{(k)}$ as per the result in the follow lemma:
\begin{lemma}\label{lem:Y-Uk}
The $k$-largest eigenvectors of $\UM^{(k)}{\UM^{(k)}}^{\top}$ is $\UM^{(k)}$.
\end{lemma} 

In a nutshell, our SCAG problem can be solved by 
(i) computing the top-$k$ left singular vectors $\YM$ of NSR $\ZM$ and (ii) converting $\YM$ into a VCA $\CM$ such that their distance is minimal.
In the rest of this section, we elaborate on the algorithmic details of \algo that implement these steps.


\eat{
which is equal to $\SM$ since $\SM=\UM^{(k)}{\UM^{(k)}}^{\top}$ is symmetric. Let $\YM\in \mathbb{R}^{n\times k}$ be a {\em vertex-cluster assignment} (VCA) matrix 
\begin{equation}
\YM_{i,j} =
\begin{cases}
\frac{1}{\sqrt{|\C_j|}} & \text{if $v_i\in \C_j$,}\\
0 & \text{Otherwise.}
\end{cases}
\end{equation}
On its basis, the objective function 
\begin{equation}
\max_{\YM^{\top}\YM=\IM}trace(\YM^{\top}\SM\YM) = trace(\YM^{\top}\UM^{(k)}{\UM^{(k)}}^{\top}\YM)
\end{equation}
}

\begin{algorithm}[!t]
\caption{\algo}\label{alg:sscag}
\KwIn{Attributed graph $\G$, the number $k$ of clusters, 
order $T$, decay factor $\alpha$, the number $\tau$ of iterations}
\KwOut{A set $\{\C_1,\C_2,\ldots,\C_k\}$ of $k$ clusters.}
\eIf{$f_{\textnormal{na\"{\i}ve}}(\G,k,\tau,T)\le f_{\textnormal{integr}}(\G,k,\tau,T)$}
{
 $\ZM \gets \itrgf{}(\hat{\PM}, \hat{\XM}, T, \alpha)$\;
 Compute $\YM^{\prime}$ by the randomized SVD with $k$ and $o$\;
}{
 Sample a Gaussian matrix $\RM \sim \mathcal{N}(0,1)^{d\times (k+o)}$\;
 \For{$t\gets 1$ \KwTo $\tau$}{
  $\HM \gets \itrgf{}(\hat{\PM}, \hat{\XM} \RM, T, \alpha)$\;
  $\HM \gets \itrgf{}(\hat{\PM}^{\top}, \HM, T, \alpha)$\;
  $\RM \gets \hat{\XM}^{\top} \HM$;\\
 }{
 $\HM \gets \itrgf{}(\hat{\PM}, \hat{\XM} \RM, T, \alpha)$\; 
 $\QM\gets$ Orthogonal matrix by a QR factorization of $\HM$\;
 $\BM \gets \itrgf{}(\hat{\PM}^{\top}, \QM, T, \alpha)$\; 
 Compute $\YM^{\prime}$ according to Eq.~\eqref{eq:comp-QY}\;
 }

}
$\{\C_1,\C_2,\ldots,\C_k\}\gets$ Invoke \textsf{SNEM} with $\YM^{\prime}_{\cdot,2:k+1}$\;
\end{algorithm}

\subsection{Algorithm}\label{sec:adap-y}
The pseudo-code of \algo is illustrated in Algo.~\ref{alg:sscag}.
\algo employs a fast rounding algorithm, \textsf{SNEM}~\cite{yang2024efficient,yang2024effective}, to fulfil the second goal, i.e., derive $\{\C_1,\C_2,\ldots,\C_k\}$ from $\YM$ (Line 14), whose runtime is merely $O(kn)$.
The critical task thus lies on the computation of $\YM$.
A naive approach proceeds as follows. First, we construct the NSR using $T$ {\em power iterations}~\cite{horn2012matrix}, dubbed as \itrgf{}\footnote{The pseudo-code appears in Appendix~\ref{sec:power-iter}.}. It takes as input $\hat{\PM}$, $\hat{\XM}$, order $T$, and decay factor $\alpha$, and outputs $\ZM$ defined in Eq.~\eqref{eq:NSR} using $O(dm)$ time per iteration.
Next, we conduct the {\em randomized SVD}~\cite{Halko2009FindingSW} of $\ZM$ to get $\YM$, which can be done in $O(kdn)$ time.
However, the cost $O(Tdm)$ of constructing $\ZM$ is significant on {\em dense} attributed graphs associated with large attribute sets.
In such cases, a better treatment is to integrate the computation of $\ZM$ into the process of the randomized SVD algorithm, so as to sidestep the explicit construction of $\ZM$.

\algo combines the aforementioned two ways in an adaptive fashion for optimal efficiency. More concretely, before entering the core procedure, \algo estimates the runtime costs of the naive and integrated approaches as the total amounts of matrix operations
\begin{small}
\begin{equation}\label{eq:cost-est}
\begin{aligned}
\textstyle f_{\text{integr}}(\G,k,\tau,T)=2(\tau+1)\cdot (k+o)\cdot(dn+Tm)+3(k+o)^2n, \\
f_{\text{na\"{\i}ve}}(\G,k,\tau,T)=Tdm+2(\tau+1)\cdot (k+o)dn +3(k+o)^2n
\end{aligned}
\end{equation}
\end{small}
therein, respectively, based on $\G$, the number $\tau$ of iterations, order $T$, and the number $k$ of clusters. 
For the sake of space, we defer the complexity analysis to Appendix~\ref{sec:algo-add}.

If $f_{\text{na\"{\i}ve}}(\G,k,\tau,T)\le f_{\text{integr}}(\G,k,\tau,T)$, Algo.~\ref{alg:sscag} follows the naive way remarked earlier (Lines 2-3), and otherwise integrates the computations of $\ZM$ and $\YM$ (Lines 5-13). To be specific, it first generates a $d\times (k+o)$ standard Gaussian random matrix $\RM$ at Line 5, where $o$ ($o\ge 2$) stands for the oversampling parameter used in randomized SVD for proper conditioning.
Next, we begin an iterative process to form $\HM\gets (\ZM\ZM^{\top})^\tau\ZM\RM$ by calling \itrgf with $\hat{\PM}$,$\hat{\XM}\RM$ or $\hat{\PM}^{\top}$, $\HM$ as inputs alternately at Lines 6-10. 

Afterwards, \algo constructs an orthonormal matrix $\QM$ through a QR factorization of $\HM$, and feeds $\hat{\PM}^{\top}$ and $\QM$ into \itrgf to get a matrix $\BM$ that builds $\QM^{\top}\ZM$ by $\BM^{\top}\hat{\XM}$ (Lines 11-12). Lastly, we calculate $\YM^{\prime}$ at Line 13 via 
\begin{equation}\label{eq:comp-QY}
\YM \gets \QM \boldsymbol{\Gamma},\ \text{where $\boldsymbol{\Gamma}$ is the left singular vectors of $\BM^{\top}\hat{\XM}$.}
\end{equation}





\begin{theorem}\label{lem:sscag}
The columns of $\YM^{\prime}$ computed at Line 13 in Algo.~\ref{alg:sscag} are the approximate top-$(k+o)$ left singular vectors of $\ZM$.
\end{theorem}

By Theorem~\ref{lem:sscag}, $\YM^{\prime}$ derived in \algo contain the approximate top-$(k+o)$ left singular vectors of $\ZM$. Particularly, \algo selects the second to $(k+1)$-th columns, i.e., $\YM^{\prime}_{\cdot,2:k+1}$, as $\YM$ for clustering. The reason is that $\ZM\ZM^{\top}$ is close to a scaled stochastic matrix, rendering $\YM^{\prime}_{\cdot,1}$ approximate ${\mathbf{1}}/{\sqrt{n}}$ that is trivial for clustering. For the interest of space, we refer readers to Appendix~\ref{sec:ZZ-stochastic} for related evidence.

\subsection{Theoretical Analysis}\label{sec:analysis}
In \algo, the VCA matrix $\CM$ is obtained via \textsf{SNEM}~\cite{yang2024efficient} with $\YM$, whose goal is to find a VCA matrix $\CM$ such that
\begin{equation*}
\min_{\TM\in\mathbb{R}^{k\times k}}{\|\YM\TM-\CM\|_2}\ \text{subject to $\TM\TM^{\top}=\IM$}
\end{equation*}
is optimized, i.e., the distance between $\YM\TM$ and $\CM$ is minimized. Ideally, the optimum $\CM^{\ast}$ satisfies $\CM^{\ast}=\YM\TM$ and $\TM\TM^{\top}$, which leads to
\begin{align*}
trace({\CM^{\ast}}^{\top}\ZM\ZM^{\top}\CM^{\ast}) & = trace(\TM^{\top}\YM^{\top}\ZM\ZM^{\top}\YM\TM) \\
& = trace(\YM^{\top}\ZM\ZM^{\top}\YM\TM\TM^{\top}) =trace(\YM^{\top}\ZM\ZM^{\top}\YM).
\end{align*}
\begin{lemma}\label{lem:Y-trace}
$\YM=\argmax{\boldsymbol{\Upsilon}\in \mathbb{R}^{n\times k}} trace(\boldsymbol{\Upsilon}^{\top}\ZM\ZM^{\top}\boldsymbol{\Upsilon})\ \text{subject to $\boldsymbol{\Upsilon}^{\top}\boldsymbol{\Upsilon}=\IM$}$.
\end{lemma}
Since $\YM$ is an optimal solution to the trace maximization problem in Lemma~\ref{lem:Y-trace}, the optimal VCA $\CM^{\ast}$ that \algo aims to derive also maximizes $trace(\CM^{\top}\ZM\ZM^{\top}\CM)$ where $\CM$ is required to be a VCA defined in Eq.~\eqref{eq:VCA}. 
Put another way, the objective of \algo is equivalent to optimizing the problem of $\max_{\CM}trace(\CM^{\top}\ZM\ZM^{\top}\CM)$.
\begin{lemma}\label{lem:conductance}
Let $\tilde{\G}$ be an undirected weighted graph with vertex set $\V$ and adjacency matrix $\WM$, wherein each entry $\WM_{i,j}$ stands for the weight of edge $(v_i,v_j)\in \tilde{\G}$. If there exists a scalar $\beta$ such that $\beta\cdot\WM$ is a stochastic matrix,
\begin{equation*}
\max_{\CM} trace(\CM^{\top}\WM^{\top}\CM) \Leftrightarrow \min_{\{\C_1,\C_2,\ldots,\C_k\}} \phi(\C_1,\C_2,\ldots,\C_k)
\end{equation*}
holds and $\phi(\C_1,\C_2,\ldots,\C_k)$ denotes the total conductance~\cite{lovasz1993random} of clusters $\{\C_1,\C_2,\ldots,\C_k\}$, i.e.,
\begin{small}
\begin{equation}\label{eq:conductance}
\textstyle \phi(\C_1,\C_2,\ldots,\C_k) = \sum_{\ell=1}^{k}\sum_{v_i\in \C_\ell,v_j\in \V\setminus \C_\ell}{\frac{\WM_{i,j}}{|\C_\ell|}}.
\end{equation}
\end{small}
\end{lemma}

Let $\tilde{\G}$ be an affinity graph constructed on the vertex set $\V$ and every edge $(v_i,v_j)\in \V\times\V$ is associated with a weight computed by $\ZM_{i}\cdot\ZM_{j}$. Accordingly, $\ZM\ZM^{\top}$ is the weighted adjacency matrix of $\tilde{\G}$, which is approximately stochastic as pinpointed in the preceding section.
Lemma~\ref{lem:conductance} implies that \algo is to partition vertices in $\tilde{\G}$ into $k$ clusters $\{\C_1,\C_2,\ldots,\C_k\}$ such that their total {\em conductance} (i.e., the overall connectivity of vertices across clusters over $\tilde{\G}$) defined in Eq.~\eqref{eq:conductance} is minimized.

\section{Modularity-based \algo Approach}\label{sec:algo-plus}
Aside from conductance, {\em modularity} introduced by~\citet{Newman2006ModularityAC} is another prominent and eminently useful metric for vertex clustering over graphs.
Inspired by our theoretical finding in Section~\ref{sec:analysis}, this section investigates incorporating the {\em modularity} maximization objective into the \algo framework and devises \algoplus for SCAG computation.

\begin{algorithm}[!t]
\caption{\algoplus}\label{alg:msscag}
\KwIn{$\G$, $k$, $T$, $\alpha$, $\gamma$, $\tau$}
\KwOut{A set of $k$ clusters $\{\C_1,\C_2,\ldots,\C_k\}$}
$\ZM \gets \itrgf{}(\hat{\PM}, \hat{\XM}, T, \alpha)$;\\
Compute $\hat{\ZM}$ and $\wvec$ by Eq.~\eqref{eq:Z-hat}\;
Sample a Gaussian matrix $\RM \sim \mathcal{N}(0,1)^{n\times k}$;\\
$\QM\gets$ Orthogonal matrix by a QR factorization of $\RM$\;
\For{$i\gets 1$ \KwTo $\tau$}{
  Compute $\HM$ according to Eq.~\eqref{eq:update-H-Q}\;
  $\QM\gets$ Orthogonal matrix by a QR factorization of $\HM$\;
}
$\{\C_1,\C_2,\ldots,\C_k\}\gets$ Invoke \textsf{SNEM} with $\QM$\;
\end{algorithm}

\subsection{High-level Idea}\label{sec:modularity}
In a fully random graph model, the probability that vertex $v_i$ is connected to $v_j$ is $\frac{d(v_i)\cdot d(v_j)}{2m}$.
Given $\G$ and a clustering $\{\C_1,\C_2,\ldots,\C_k\}$, the modularity~\cite{Newman2006ModularityAC} of $\{\C_1,\C_2,\ldots,\C_k\}$ on $\G$ measures the deviation of the intra-cluster connectivity on $\G$ from what would be observed in expectation when edges in $\G$ are randomly populated:
\begin{small}
\begin{equation*}
 Q =\frac{1}{2m}\sum_{\ell=1}^{k}\sum_{v_i,v_j\in \C_\ell}{ \AM_{i,j} - \frac{d(v_i)\cdot d(v_j)}{2m}}.
\end{equation*}
\end{small}
Modularity-based clustering aims to find a division $\{\C_1,\C_2,\ldots,\C_k\}$ of $\G$ that maximizes modularity $Q$.

Akin to the conductance minimization objective in \algo (Section~\ref{sec:analysis}), we extend the modularity definition to the weighted graph $\tilde{\G}$ constructed based on NSR $\ZM$ and formulate $Q$ over $\tilde{\G}$ as follows:
\begin{small}
\begin{equation}\label{eq:modularity-W}
\textstyle Q =\frac{1}{{\wvec^{\top}\boldsymbol{1}}}\sum_{\ell=1}^{k}\sum_{v_i,v_j\in \C_\ell}{ \WM_{i,j} - \gamma\cdot \frac{\wvec_i\cdot \wvec_j}{{\wvec^{\top}\boldsymbol{1}}}},
\end{equation}
\end{small}
where $\gamma$ is a weight parameter for the second term.
$\WM_{i,j}$ denotes the weight of each edge $(v_i,v_j)\in \V\times\V$
\begin{small}
\begin{equation}\label{eq:Z-hat}
\textstyle \WM_{i,j} = \hat{\ZM}_i\cdot {\hat{\ZM}_j}^{\top}\ \text{where}\ \hat{\ZM}_i = \frac{\ZM_i}{{\sqrt{\sum_{v_j\in \V}\ZM_i\cdot {\ZM_j}^{\top}}}}\ \forall{v_i\in \V},
\end{equation}
\end{small}
and $\wvec\in \mathbb{R}^n$ is a length-$n$ column vector where each entry $\wvec_i=\sum_{v_j\in \V}{\WM_{i,j}}\ \forall{v_i\in \V}$. Unlike \algo, we leverage the normalized version $\hat{\ZM}$ of $\ZM$ for edge weighting in affinity graph $\tilde{\G}$ as $\ZM\ZM^{\top}$ is approximately stochastic as pinpointed previously, which makes $\wvec$ an all-one vector and thus the term $\gamma \frac{\wvec_i\cdot \wvec_j}{{\wvec^{\top}\boldsymbol{1}}}$ in $Q$ ineffectual. 



\begin{lemma}\label{lem:mod-trace}
Given $k$ clusters $\{\C_1,\C_2,\ldots,\C_k\}$, whose corresponding vertex-cluster indicator is $\CM \in \mathbb{R}^{n\times k}$ where $\CM_{i,j}=1$ if $v_i\in \C_j$ and $0$ otherwise. If the following trace maximization problem is optimized
\begin{small}
\begin{equation}\label{eq:mod-trace-max}
\textstyle \max_{\CM} trace\left(\CM^{\top}\left(\hat{\ZM}\hat{\ZM}^{\top} - \gamma \cdot \frac{\wvec\wvec^{\top}}{\wvec^{\top}\boldsymbol{1}} \right)\CM \right),
\end{equation}
\end{small}
$\{\C_1,\C_2,\ldots,\C_k\}$ maximizes the modularity $Q$ defined on $\tilde{\G}$ in Eq.~\eqref{eq:modularity-W}.
\end{lemma}

Lemma~\ref{lem:mod-trace} establishes an equivalence between our modularity-based objective in Eq.~\eqref{eq:modularity-W} and the trace maximization problem in Eq.~\eqref{eq:mod-trace-max}.
By relaxing $\CM$ and leveraging Ky Fan’s trace maximization principle~\cite{Fan1949OnAT}, the matrix $\CM$ optimizing Eq.~\eqref{eq:mod-trace-max} is the $k$-largest eigenvectors $\YM$ of $\hat{\ZM}\hat{\ZM}^{\top} - \gamma \cdot \frac{\wvec\wvec^{\top}}{\wvec^{\top}\boldsymbol{1}}$. 
Analogous to \algo, the clusters $\{\C_1,\C_2,\ldots,\C_k\}$ thus can be extracted from $\YM$ through \textsf{SNEM}. 

Instead of materializing the $n\times n$ dense matrix $\hat{\ZM}\hat{\ZM}^{\top} - \gamma \cdot \frac{\wvec\wvec^{\top}}{\wvec^{\top}\boldsymbol{1}}$ for the computation of $\YM$, the idea of \algoplus is to harness the fact that the matrix-vector product $(\hat{\ZM}\hat{\ZM}^{\top} - \gamma \cdot \frac{\wvec\wvec^{\top}}{\wvec^{\top}\boldsymbol{1}})\cdot \qvec$ in iterative eigenvalue solvers can be reordered as $\hat{\ZM}\cdot(\hat{\ZM}^{\top}\qvec) - \gamma \cdot \frac{\wvec}{\wvec^{\top}\boldsymbol{1}}\cdot(\wvec^{\top}\qvec)$ and done in $O(dn)$ time.

\subsection{Algorithm and Analysis}
Algo.~\ref{alg:msscag} presents the pseudo-code of \algoplus, which begins by taking an additional parameter $\gamma$ compared to Algo.~\ref{alg:sscag}. Initially, \algoplus obtains NSR $\ZM$ via \itrgf, based on which it calculates $\hat{\ZM}$ and $\wvec$ as in the context of Eq.~\eqref{eq:Z-hat} (Lines 1-2). At Lines 3-4, we create an $n\times k$ orthogonal matrix $\QM$ through a QR decomposition of a standard Gaussian matrix $\RM\in \mathbb{R}^{n\times k}$ generated randomly. Subsequently, \algoplus{} performs $\tau$ {\em subspace iterations}~\cite{saad2011numerical} to update $\QM$, each of which first calculates $\HM$ at Line 6 by
\begin{small}
\begin{equation}\label{eq:update-H-Q}
\textstyle \HM \gets \hat{\ZM}\cdot \left(\hat{\ZM}^{\top} \QM\right)-\gamma\cdot\frac{\wvec}{\wvec^{\top}\mathbf{1}}\cdot \left(\wvec^{\top}\QM\right),
\end{equation}
\end{small}
and then at Line 7 updates $\QM$ as the orthogonal matrix from the QR factorization of $\HM$. The final $\QM$ will be used as $\YM$ input to \textsf{SNEM} for outputting clusters $\{\C_1,\C_2,\ldots,\C_k\}$.

\begin{theorem}\label{lem:m-sscag}
When $\QM$ in Algo.~\ref{alg:msscag} converges, the columns of $\QM$ are the $k$-largest eigenvectors of $\hat{\ZM}\hat{\ZM}^{\top} - \gamma \cdot \frac{\wvec\wvec^{\top}}{\wvec^{\top}\boldsymbol{1}}$ and $\SM=\QM\QM^{\top}$ optimizes
\begin{small}
\begin{equation}\label{eq:obj-M}
\min_{rank(\SM)=k}\|\tilde{\ZM}-\SM\tilde{\ZM}\|_F^2 + \|\SM^{\top}\SM-\IM\|_F^2,
\end{equation}
\end{small}
where $\tilde{\ZM}$ satisfies $\textstyle \tilde{\ZM}\tilde{\ZM}^{\top} = \hat{\ZM}\hat{\ZM}^{\top} - \gamma \cdot \frac{\wvec\wvec^{\top}}{\wvec^{\top}\boldsymbol{1}}$.
\end{theorem}

Theorem~\ref{lem:m-sscag} proves the correctness of \algoplus and manifests that the objective function of \algoplus from the perspective of subspace clustering can be formulated as Eq.~\eqref{eq:obj-M} based on NSR $\tilde{\ZM}$ ensuring $\textstyle \tilde{\ZM}\tilde{\ZM}^{\top} = \hat{\ZM}\hat{\ZM}^{\top} - \gamma \cdot \frac{\wvec\wvec^{\top}}{\wvec^{\top}\boldsymbol{1}}$.

\section{Experiments}
This section experimentally evaluates \algo and \algoplus against 17 competitors regarding clustering quality and efficiency on 8 real attributed graphs.
All experiments are conducted on a Linux machine with an NVIDIA Ampere A100 GPU (80GB memory), AMD EPYC 7513 CPUs (2.6 GHz), and 1TB RAM.
Due to space constraint, we defer the clustering visualizations to Appendix~\ref{sec:add-exp}.

\begin{table}[!t]
\centering
\renewcommand{\arraystretch}{0.8}
\begin{small}
\caption{Statistics of Datasets.}\label{tbl:exp-data}
\vspace{-3mm}
\begin{tabular}{l|r|r|r|c}
	\hline
	{\bf Dataset} & \multicolumn{1}{c|}{\bf \#Vertices } & \multicolumn{1}{c|}{\bf \#Edges } & \multicolumn{1}{c|}{\bf \#Attributes } & \multicolumn{1}{c}{\bf \#Clusters}  \\
	\hline
    { {\em CiteSeer}} & 3,327 & 4,732 & 3,703 & 6   \\
    { {\em Wiki}} & 2,405 & 14,001 & 4,973 & 17  \\
    { {\em ACM}} & 3,025 & 16,153 & 1,870  & 3 \\
    { {\em Photo}} & 7,487 & 119,043 & 745 & 8   \\
    { {\em Cora}} & 19,793 & 63,421 & 8,710 & 70   \\
    { {\em PubMed}}  & 19,717 & 64,041 & 500 & 3   \\
    { {\em DBLP}}  & 4,057 & 2,502,276 & 334 & 4  \\
    { {\em ArXiv}} & 169,343 & 1,327,142 & 128 & 40   \\
    \hline
\end{tabular}%
\end{small}
\vspace{-2ex}
\end{table}

\subsection{Experimental Setup}\label{sec:exp-setup}
\stitle{Datasets}
Table~\ref{tbl:exp-data} summarizes the statistics of datasets we experiment with. {\em CiteSeer}, {\em PubMed}~\cite{sen2008cocl}, {\em Cora}~\cite{Bojchevski2017DeepGE}, {\em ACM}, {\em DBLP}~\cite{han2019}, and {\em ArXiv}~\cite{hu2020ogb} are academic citation networks, in which 
ground-truth clusters represent subjects or fields of study of publications.
{\em Photo} is a segment of the Amazon product co-purchase graph~\cite{MuAley2015Ima}, where
cluster labels correspond to product categories.
{\em Wiki}~\cite{Yang2015Net} is a reference network of Wikipedia documents.

\stitle{Evaluation Criteria}
Following previous works~\cite{fettal2023scalable,liu2023dink,tsitsulin2023graph,Cai2022EfficientDE,bhowmick2024dgcluster}, we adopt three widely-used metrics: {\em Clustering Accuracy} (ACC), {\em Normalized Mutual Information} (NMI), and {\em Adjusted Rand Index} (ARI) to assess the clustering quality in the presence of ground-truth cluster labels. ACC and NMI scores range from $0$ to $1.0$, whilst ARI ranges from $-0.5$ to $1.0$. For all of them, higher values indicate better clustering performance.

\begin{table*}[!t]
\centering
\renewcommand{\arraystretch}{0.8}
\caption{Clustering Quality on {\em ACM}, {\em Wiki}, {\em CiteSeer}, and {\em Photo}.}\vspace{-3mm}
\begin{small}
\addtolength{\tabcolsep}{-0.25em}
\begin{tabular}{c|ccc | ccc | ccc | ccc|c  }
\hline
\multirow{2}{*}{\bf Method} & \multicolumn{3}{c|}{\bf{ {\em ACM}}} & \multicolumn{3}{c|}{\bf{ {\em Wiki}}}   & \multicolumn{3}{c|}{\bf{ {\em CiteSeer}}} & \multicolumn{3}{c|}{\bf{ {\em Photo}}}  & \multirow{2}{*}{\bf Rank}\\ \cline{2-13}
& ACC \textuparrow & NMI \textuparrow & ARI \textuparrow & ACC \textuparrow & NMI \textuparrow & ARI \textuparrow & ACC \textuparrow & NMI \textuparrow & ARI \textuparrow & ACC \textuparrow & NMI \textuparrow & ARI \textuparrow  \\ 
\hline
\textsf{KMeans} \cite{Ketchen1996kmeans} 	&$88.7_{\pm 0.9}$	&$63.3_{\pm 1.7}$&$69.4_{\pm 2.2}$&$45.6_{\pm 2.9}$&$46.9_{\pm 2.1}$&$26.1_{\pm 2.5}$&$29.84_{\pm 1.8}$	&$39.8_{\pm 0.5}$&$36.8_{\pm 0.7}$&$43.8_{\pm 1.6}$&$34.3_{\pm 0.8}$&$23.6_{\pm 1.0}$&$11.2$\\ \hline
\textsf{K-FSC} \cite{Fan2021kfsc} 	&$70.5_{\pm 0.0}$&$28.1_{\pm 0.0}$&$31.3_{\pm 0.0}$&$47.2_{\pm 1.2}$&$44.1_{\pm 1.9}$&$27.8_{\pm 1.8}$&$18.5_{\pm 0.0}$	&$0.0_{\pm 0.0}$	&$0.0_{\pm 0.0}$	&$47.7_{\pm 0.0}$		& $23.3_{\pm 0.0}$ & $18.5_{\pm 0.0}$&$13.6$		 	    \\
\textsf{LSR} \cite{Lu2012Lsr} 	&$60.3_{\pm 0.0}$&$2.1_{\pm 0.0}$&$3.7_{\pm 0.0}$&$7.4_{\pm 0.0}$&$2.5_{\pm 0.0}$&$0.1_{\pm 0.0}$	&$18.5_{\pm 3.1}$	&$0.0_{\pm 0.0}$&$0.0_{\pm 0.0}$		&$3.2_{\pm 0.0}$ &$33.5_{\pm 0.0}$&$12.7_{\pm 0.0}$ &$17.5$  \\ 
\textsf{SSC-OMP} \cite{You2015ScalableSS} 	&$81.5_{\pm 0.0}$&$47.6_{\pm 0.0}$&$53.6_{\pm 0.0}$&$42.6_{\pm 2.1}$&$38.4_{\pm 0.9}$&$24.4_{\pm 1.2}$&$22.7_{\pm 3.1}$&$2.7_{\pm 3.1}$&$1.3_{\pm 2.1}$&$60.0_{\pm 0.1}$&$49.9_{\pm 0.4}$&$40.9_{\pm 0.3}$&$14.5$  \\ 
\textsf{EDESC} \cite{Cai2022EfficientDE} 	&$82.5_{\pm 0.0}$&$52.9_{\pm 0.0}$&$56.1_{\pm 0.0}$&$40.9_{\pm 0.0}$&$37.5_{\pm 0.0}$&$20.2_{\pm 0.0}$&$42.9_{\pm 0.0}$&$19.8_{\pm 0.0}$&$16.0_{\pm 0.0}$&$38.8_{\pm 0.0}$&$26.8_{\pm 0.0}$&$16.8_{\pm 0.0}$&$15.3$\\
\textsf{SAGSC} \cite{fettal2023scalable} 	& \underline{$93.2_{\pm 0.0}$} &\underline{$75.1_{\pm 0.1}$}&\underline{$80.9_{\pm 0.1}$}&$55.1_{\pm 2.2}$&$52.5_{\pm 1.0}$&$32.5_{\pm 2.6}$&$67.7_{\pm 0.0}$&$42.9_{\pm 0.1}$&$43.5_{\pm 0.1}$&\underline{$77.9_{\pm 0.0}$}&\cellcolor{blue!30}\underline{$71.9_{\pm 0.0}$}&\cellcolor{blue!30}\underline{$60.2_{\pm 0.0}$}& \underline{$3.8$}\\ \hline
\textsf{SC} \cite{shi2000normalized} 	&$84.5_{\pm 0.0}$&$53.7_{\pm 0.1}$&$59.7_{\pm 0.0}$&$19.9_{\pm 2.1}$&$8.5_{\pm 2.4}$&$0.2_{\pm 0.2}$&$20.9_{\pm 0.0}$&$1.4_{\pm 0.0}$&$0.0_{\pm 0.0}$&$75.1_{\pm 0.0}$&$59.7_{\pm 0.0}$&$54.4_{\pm 0.0}$&$15.0$\\
\textsf{MinCutPool} \cite{bianchi2020mincutpool} 	&$89.8_{\pm 0.0}$&$65.6_{\pm 0.0}$&$71.8_{\pm 0.0}$&$44.0_{\pm 0.0}$&$38.8_{\pm 0.0}$&$24.1_{\pm 0.0}$&$40.4_{\pm 0.0}$&$21.5_{\pm 0.0}$&$16.7_{\pm 0.0}$&$25.4_{\pm 0.0}$&$0.2_{\pm 0.0}$&$0.0_{\pm 0.0}$&$11.6$\\
\textsf{DMoN} \cite{tsitsulin2023graph} 	&$34.5_{\pm 0.1}$&$0.4_{\pm 0.0}$&$0.1_{\pm 0.0}$&$52.2_{\pm 0.0}$&$47.4_{\pm 0.0}$&$32.2_{\pm 0.1}$&$30.6_{\pm 0.1}$&$26.5_{\pm 0.0}$&$16.8_{\pm 0.0}$&$23.9_{\pm 0.3}$&$9.8_{\pm 0.3}$&$4.1_{\pm 0.2}$&$13.0$\\
\textsf{DGCluster} \cite{bhowmick2024dgcluster} 	&$68.2_{\pm 0.0}$&$62.9_{\pm 0.0}$&$58.2_{\pm 0.0}$&\underline{$56.4_{\pm 0.0}$}&$50.2_{\pm 0.0}$&\cellcolor{blue!15}\underline{$40.6_{\pm 0.0}$}&$39.8_{\pm 0.0}$&$41.1_{\pm 0.0}$&$27.1_{\pm 0.0}$&$72.0_{\pm 0.0}$&$70.7_{\pm 0.0}$&$58.0_{\pm 0.0}$&$7.2$\\ \hline
\textsf{GCC} \cite{fettal2022efficient} 	&$65.3_{\pm 0.0}$&$55.0_{\pm 0.1}$&$48.0_{\pm 0.0}$&$54.8_{\pm 0.0}$&\cellcolor{blue!15}\underline{$55.1_{\pm 0.0}$}&$33.8_{\pm 0.0}$&$69.4_{\pm 0.0}$&$45.1_{\pm 0.0}$&$45.5_{\pm 0.0}$&$59.7_{\pm 2.2}$&$61.1_{\pm 5.0}$&$38.7_{\pm 1.7}$&$6.2$\\
\textsf{GIC}~\cite{Mavromatis2021GraphIM}   	&$90.7_{\pm 0.0}$&$70.7_{\pm 0.0}$&$73.5_{\pm 0.0}$&$44.2_{\pm 0.0}$&$44.7_{\pm 0.0}$&$28.3_{\pm 0.0}$&$68.3_{\pm 0.0}$&$44.5_{\pm 0.0}$&$46.0_{\pm 0.0}$&$65.6_{\pm 0.1}$&$59.7_{\pm 0.0}$&$47.9_{\pm 0.0}$&$8.3$\\ 
\textsf{SSGC} \cite{Zhu2021SimpleSG} 	&$86.9_{\pm 0.0}$&$61.2_{\pm 0.0}$&$65.7_{\pm 0.0}$&$50.5_{\pm 0.0}$&$47.7_{\pm 0.0}$&$27.4_{\pm 0.0}$&$69.0_{\pm 0.0}$&$42.8_{\pm 0.0}$&$44.4_{\pm 0.0}$&$57.7_{\pm 0.1}$&$61.2_{\pm 0.0}$&$33.8_{\pm 0.0}$&$7.8$\\
\textsf{SDCN} \cite{sdcn2020} 	&$89.8_{\pm 0.0}$&$66.4_{\pm 0.0}$&$72.2_{\pm 0.0}$&$36.5_{\pm 0.0}$&$31.7_{\pm 0.0}$&$16.5_{\pm 0.0}$&$47.0_{\pm 0.1}$&$23.4_{\pm 0.0}$&$18.7_{\pm 0.1}$&$53.4_{\pm 0.1}$&$40.8_{\pm 0.1}$&$31.8_{\pm 0.1}$&$11.9$\\ 
\textsf{DCRN} \cite{DCRN} 	&$89.3_{\pm 0.0}$&$65.5_{\pm 0.0}$&$70.8_{\pm 0.0}$&$51.6_{\pm 0.0}$&$49.4_{\pm 0.0}$&$22.4_{\pm 0.0}$&\cellcolor{blue!15} \underline{$70.6_{\pm 0.0}$}&\cellcolor{blue!15}\underline{$45.6_{\pm 0.0}$}&\underline{$47.6_{\pm 0.1}$}&$75.7_{\pm 0.1}$&$70.7_{\pm 0.1}$&$57.3_{\pm 0.1}$&$6.5$\\
\textsf{DAEGC} \cite{wang2019attributed} 	&$90.1_{\pm 0.0}$&$67.6_{\pm 0.0}$&$73.1_{\pm 0.0}$&$45.7_{\pm 0.0}$&$41.9_{\pm 0.0}$&$25.8_{\pm 0.0}$&$68.4_{\pm 0.0}$&$43.9_{\pm 0.0}$&$44.4_{\pm 0.0}$&$41.4_{\pm 0.0}$&$36.5_{\pm 0.0}$&$13.4_{\pm 0.0}$&$9.1$\\
\textsf{Dink-Net} \cite{liu2023dink} 	&$62.2_{\pm 0.0}$&$27.1_{\pm 0.0}$&$23.2_{\pm 0.0}$&$42.9_{\pm 0.0}$&$35.7_{\pm 0.0}$&$21.3_{\pm 0.0}$&$67.6_{\pm 0.0}$&$32.4_{\pm 0.0}$&$32.1_{\pm 0.0}$&$71.4_{\pm 0.0}$&$59.7_{\pm 0.0}$&$49.7_{\pm 0.0}$&$11.7$\\ \hline
\algo (\textsf{SNEM})  	&\cellcolor{blue!15} $93.5_{\pm 0.0}$ &\cellcolor{blue!15}  $75.4_{\pm 0.0}$& \cellcolor{blue!15} $81.4_{\pm 0.0}$&\cellcolor{blue!30}$64.4_{\pm 0.0}$&\cellcolor{blue!30}$55.1_{\pm 0.0}$&\cellcolor{blue!30}$44.9_{\pm 0.0}$&\cellcolor{blue!30}$72.2_{\pm 0.0}$&\cellcolor{blue!30}$45.6_{\pm 0.0}$&\cellcolor{blue!30}$48.5_{\pm 0.0}$&\cellcolor{blue!15}$78.9_{\pm 0.0}$
&$69.0_{\pm 0.0}$&$58.9_{\pm 0.1}$& \cellcolor{blue!15}$2.0$	 	    \\ 
\algoplus (\textsf{SNEM}) 	&\cellcolor{blue!30}$93.7_{\pm 0.0}$
 &\cellcolor{blue!30}$76.0_{\pm 0.0}$&\cellcolor{blue!30}$82.0_{\pm 0.0}$&\cellcolor{blue!15}$60.2_{\pm 0.0}$&$51.1_{\pm 0.0}$&$37.8_{\pm 0.0}$&\cellcolor{blue!30}$72.2_{\pm 0.0}$&$45.2_{\pm 0.0}$&\cellcolor{blue!15}$48.2_{\pm 0.0}$&\cellcolor{blue!30}$79.2_{\pm 0.0}$&\cellcolor{blue!15}$71.1_{\pm 0.0}$&\cellcolor{blue!15}$59.7_{\pm 0.0}$& \cellcolor{blue!30}$1.7$\\ \hline
\end{tabular}
\end{small}
\label{tbl:node-clustering-small}
\vspace{0ex}
\end{table*}

\begin{table*}[!t]
\centering
\renewcommand{\arraystretch}{0.8}
\caption{Clustering Quality on {\em DBLP}, {\em PubMed}, {\em Cora}, and {\em ArXiv}.}\vspace{-3mm}
\begin{small}
\addtolength{\tabcolsep}{-0.25em}
\begin{tabular}{c|ccc | ccc | ccc | ccc |c  }
\hline
\multirow{2}{*}{\bf Method} &\multicolumn{3}{c|}{\bf{ {\em DBLP}}}& \multicolumn{3}{c|}{\bf{ {\em PubMed}}} & \multicolumn{3}{c|}{\bf{ {\em Cora}}}  & \multicolumn{3}{c|}{\bf{ {\em ArXiv}}}  & \multirow{2}{*}{\bf Rank}\\ \cline{2-13}
& ACC \textuparrow & NMI \textuparrow & ARI \textuparrow & ACC \textuparrow & NMI \textuparrow & ARI \textuparrow & ACC \textuparrow & NMI \textuparrow & ARI \textuparrow & ACC \textuparrow & NMI \textuparrow & ARI \textuparrow   \\ 
\hline
\textsf{KMeans} \cite{Ketchen1996kmeans} 	&$68.1_{\pm 0.2}$&$37.3_{\pm 0.2}$&$31.87_{\pm 0.5}$&$59.2_{\pm 0.1}$&$30.0_{\pm 0.1}$&$26.6_{\pm 0.1}$&$29.8_{\pm 1.8}$&$39.8_{\pm 0.5}$&$10.5_{\pm 0.4}$&$17.0_{\pm 0.2}$&$22.6_{\pm 0.1}$&$7.3_{\pm 0.1}$&$11.2$\\ \hline
\textsf{K-FSC} \cite{Fan2021kfsc} 	&$42.0_{\pm 0.0}$	&$13.2_{\pm 0.0}$	&$10.3_{\pm 0.0}$&$55.4_{\pm 0.1}$&$16.5_{\pm 0.0}$&$15.7_{\pm 0.0}$		&$23.9_{\pm 0.0}$&$29.8_{\pm 0.0}$&$11.0_{\pm 0.0}$	&$17.3_{\pm 0.0}$&$15.5_{\pm 0.0}$&$6.1_{\pm 0.0}$&$13.6$	 	    \\
\textsf{LSR} \cite{Lu2012Lsr} 	&$30.9_{\pm 0.2}$&$1.1_{\pm 0.1}$&$0.3_{\pm 0.1}$&$38.8_{\pm 0.0}$&$0.4_{\pm 0.0}$&$0.6_{\pm 0.0}$&$1.8_{\pm 0.0}$&$0.1_{\pm 0.0}$&$0.0_{\pm 0.0}$&$6.3_{\pm 0.0}$&$0.5_{\pm 0.0}$&$0.0_{\pm 0.0}$&$17.5$  \\ 
\textsf{SSC-OMP} \cite{You2015ScalableSS} 	&$29.5_{\pm 0.1}$&$0.2_{\pm 0.1}$&$-0.1_{\pm 0.0}$&$60.3_{\pm 0.0}$&$21.5_{\pm 0.0}$&$18.9_{\pm 0.0}$&$13.8_{\pm 0.2}$&$22.6_{\pm 0.1}$&$6.3_{\pm 0.1}$	&-&-&-&$14.5$	     \\ 
\textsf{EDESC} \cite{Cai2022EfficientDE} 	&$44.5_{\pm 0.1}$&$12.8_{\pm 0.1}$&$12.3_{\pm 0.1}$&$47.8_{\pm 0.0}$&$3.8_{\pm 0.0}$&$3.1_{\pm 0.0}$&$10.5_{\pm 0.0}$&$15.2_{\pm 0.0}$&$2.9_{\pm 0.0}$	&-&-&-&$15.3$	    \\
\textsf{SAGSC} \cite{fettal2023scalable} 	&\cellcolor{blue!15} \underline{$93.1_{\pm 0.1}$}&\underline{$78.0_{\pm 0.2}$}& $83.3_{\pm 0.3}$&\underline{$71.1_{\pm 0.0}$}&$32.9_{\pm 0.0}$&\underline{$34.1_{\pm 0.0}$}&\underline{$41.3_{\pm 0.9}$}&$53.9_{\pm 0.9}$&$24.4_{\pm 1.3}$	&\cellcolor{blue!15}\underline{$47.8_{\pm 1.8}$}&$46.9_{\pm 0.6}$&\underline{$38.2_{\pm 0.7}$}& \underline{$3.8$}	     \\ \hline
\textsf{SC} \cite{shi2000normalized} 	&$30.0_{\pm 0.0}$&$1.4_{\pm 0.0}$&$0.1_{\pm 0.0}$&$53.8_{\pm 0.0}$&$10.8_{\pm 0.0}$&$7.2_{\pm 0.0}$&$10.2_{\pm 0.5}$&$16.3_{\pm 1.0}$&$2.0_{\pm 0.2}$	&-&-&-	&$15.0$     \\
\textsf{MiniCutPool} \cite{bianchi2020mincutpool} 	&$87.7_{\pm 0.1}$&$71.8_{\pm 0.0}$&$74.1_{\pm 0.0}$&$61.5_{\pm 0.0}$&$28.3_{\pm 0.0}$&$25.4_{\pm 0.0}$&$31.0_{\pm 0.0}$&$49.5_{\pm 0.0}$&$19.1_{\pm 0.0}$	&-&-&-	 &$11.6$    \\
\textsf{DMoN} \cite{tsitsulin2023graph} 	&$46.3_{\pm 0.6}$&$57.3_{\pm 0.1}$&$38.4_{\pm 0.5}$&$21.7_{\pm 0.0}$&$22.6_{\pm 0.0}$ &$9.5_{\pm 0.0}$&$23.3_{\pm 0.0}$&$42.7_{\pm 0.0}$ &$16.4_{\pm 0.0}$&$37.5_{\pm 0.0}$&$38.6_{\pm 0.0}$&$27.4_{\pm 0.0}$ &$13.0$\\
\textsf{DGCluster} \cite{bhowmick2024dgcluster} 	&$92.1_{\pm 0.1}$&$75.2_{\pm 0.0}$&$80.9_{\pm 0.0}$&$41.4_{\pm 0.0}$&\underline{$34.7_{\pm 0.0}$}&$24.4_{\pm 0.0}$&$38.1_{\pm 0.0}$&$53.4_{\pm 0.0}$&$28.0_{\pm 0.0}$	&$33.9_{\pm 0.0}$&$43.2_{\pm 0.0}$&$29.2_{\pm 0.0}$&$7.2$\\ \hline
\textsf{GCC} \cite{fettal2022efficient} 	&$90.9_{\pm 1.1}$&$73.6_{\pm 0.0}$&$78.4_{\pm 2.2}$&$70.8_{\pm 2.2}$&$32.3_{\pm 4.1}$&$33.3_{\pm 0.0}$&$40.2_{\pm 0.0}$&\cellcolor{blue!30}\underline{$54.1_{\pm 0.0}$}&$26.0_{\pm 0.0}$	&$41.2_{\pm 0.0}$&\cellcolor{blue!15}\underline{$47.1_{\pm 0.0}$}&$35.8_{\pm 0.0}$&$6.2$\\
\textsf{GIC}~\cite{Mavromatis2021GraphIM}   	&$87.4_{\pm 0.0}$&$68.1_{\pm 0.0}$&$71.7_{\pm 0.0}$&$65.1_{\pm 0.0}$&$26.4_{\pm 0.0}$&$24.6_{\pm 0.0}$&$30.6_{\pm 0.0}$&$50.2_{\pm 0.0}$&$20.0_{\pm 0.0}$	&$12.4_{\pm 0.0}$&$18.4_{\pm 0.0}$&$6.4_{\pm 0.0}$&$8.3$	 \\ 
\textsf{SSGC} \cite{Zhu2021SimpleSG} 	&$87.1_{\pm 0.0}$&$67.5_{\pm 0.0}$&$71.6_{\pm 0.0}$&$70.7_{\pm 0.0}$&$32.5_{\pm 4.5}$&$33.3_{\pm 0.0}$&$37.0_{\pm 0.0}$&$53.6_{\pm 0.0}$&$26.0_{\pm 0.0}$	&$28.9_{\pm 0.0}$&$32.0_{\pm 0.0}$&$20.0_{\pm 0.0}$&$7.8$\\
\textsf{SDCN} \cite{sdcn2020} 	&$79.4_{\pm 0.1}$&$58.3_{\pm 0.0}$&$58.7_{\pm 0.1}$&$63.0_{\pm 0.0}$&$23.4_{\pm 0.0}$&$22.2_{\pm 0.0}$&$11.5_{\pm 0.0}$&$21.1_{\pm 0.0}$&$4.2_{\pm 0.0}$	&$26.4_{\pm 0.0}$&$17.8_{\pm 0.0}$&$10.5_{\pm 0.0}$&$11.9$	      \\ 
\textsf{DCRN} \cite{DCRN} 	&$93.0_{\pm 0.0}$&$77.8_{\pm 0.0}$&\cellcolor{blue!15}\underline{$83.5_{\pm 0.0}$}&$69.0_{\pm 0.0}$&$34.2_{\pm 0.0}$&$32.1_{\pm 0.0}$&$39.4_{\pm 0.0}$&$53.3_{\pm 0.0}$&\underline{$26.5_{\pm 0.0}$}&-&-&- &$6.5$   \\
\textsf{DAEGC} \cite{wang2019attributed} 	&$89.8_{\pm 0.0}$&$71.2_{\pm 0.0}$&$76.1_{\pm 0.0}$&$65.2_{\pm 0.0}$&$25.8_{\pm 0.0}$&$24.6_{\pm 0.0}$&$32.2_{\pm 0.0}$&$49.9_{\pm 0.0}$&$21.3_{\pm 0.0}$&-&-&- &$9.1$   \\
\textsf{Dink-Net} \cite{liu2023dink} 	&$87.3_{\pm 0.1}$&$67.1_{\pm 0.0}$&$69.7_{\pm 0.0}$&$66.1_{\pm 0.0}$&$25.7_{\pm 0.0}$&$25.9_{\pm 0.0}$&$32.9_{\pm 0.0}$&$39.3_{\pm 0.0}$&$15.8_{\pm 0.0}$	&$20.5_{\pm 0.0}$&$17.6_{\pm 0.0}$&$7.1_{\pm 0.0}$&$11.7$\\ \hline
\algo (\textsf{SNEM}) &\cellcolor{blue!30}$93.5_{\pm 0.0}$&\cellcolor{blue!15}$78.8_{\pm 0.0}$&\cellcolor{blue!30}$84.3_{\pm 0.0}$&\cellcolor{blue!15}$75.3_{\pm 0.0}$&\cellcolor{blue!15}$36.5_{\pm 0.0}$&\cellcolor{blue!15}$41.9_{\pm 0.0}$&\cellcolor{blue!30}$44.7_{\pm 0.0}$&$53.6_{\pm 0.0}$&\cellcolor{blue!30}$33.1_{\pm 0.0}$&$46.9_{\pm 0.0}$&$46.1_{\pm 0.0}$&\cellcolor{blue!15} $38.7_{\pm 0.0}$& \cellcolor{blue!15}$2.0$\\ 
\algoplus (\textsf{SNEM}) 	&\cellcolor{blue!30}$93.5_{\pm 0.0}$&\cellcolor{blue!30}$78.9_{\pm 0.0}$&\cellcolor{blue!30}$84.3_{\pm 0.0}$&\cellcolor{blue!30}$75.5_{\pm 0.0}$&\cellcolor{blue!30}$36.8_{\pm 0.0}$&\cellcolor{blue!30}$42.3_{\pm 0.0}$&\cellcolor{blue!15}$44.4_{\pm 0.0}$&\cellcolor{blue!15}$53.9_{\pm 0.0}$&\cellcolor{blue!15}$31.6_{\pm 0.0}$&\cellcolor{blue!30}$50.0_{\pm 0.0}$&\cellcolor{blue!30}$47.2_{\pm 0.0}$&\cellcolor{blue!30}$40.5_{\pm 0.0}$& \cellcolor{blue!30}$1.7$\\ \hline

\end{tabular}
\end{small}
\label{tbl:node-clustering-large}
\vspace{0ex}
\end{table*}

\stitle{Baselines, Implementations, and Parameter Settings}
We carefully select 17 competing methods from four categories for comparison including
one metric clustering method \textsf{KMeans} \cite{Ketchen1996kmeans};
five subspace clustering methods: \textsf{K-FSC} \cite{Fan2021kfsc}, \textsf{LSR} \cite{Lu2012Lsr}, \textsf{SSC-OMP} \cite{You2015ScalableSS}, \textsf{EDESC} \cite{Cai2022EfficientDE}, \textsf{SAGSC} \cite{fettal2023scalable};
four spectral methods: \textsf{SC} \cite{shi2000normalized}, \textsf{MinCutPool} \cite{bianchi2020mincutpool}, \textsf{DMoN} \cite{tsitsulin2023graph}, \textsf{DGCluster} \cite{bhowmick2024dgcluster};
and seven GRL-based methods: \textsf{GCC} \cite{fettal2022efficient}, \textsf{GIC}~\cite{Mavromatis2021GraphIM}, \textsf{SSGC} \cite{Zhu2021SimpleSG}, \textsf{SDCN} \cite{sdcn2020}, \textsf{DCRN} \cite{DCRN}, \textsf{DAEGC} \cite{wang2019attributed}, \textsf{Dink-Net} \cite{liu2023dink}.
Amid them, metric clustering and subspace clustering methods are solely applied to attribute matrices, except \textsf{SAGSC}, which is a SCAG solution using hand-crafted features from graph structures and attributes.
Spectral methods produce clusters by optimizing conductance- or modularity-like metrics on the original graph,
while GRL-based approaches apply \textsf{KMeans} to vertex representations learned by various neural network models.

For most competitors, we reproduce results using source codes collected from authors and parameter settings prescribed when possible.
Unless otherwise specified, we set $\gamma$ in \algoplus to $0.9$ on {\em CiteSeer} and {\em Wiki} and $1.0$ on others. 
As for the numbers $\tau$ of iterations in \algo and \algoplus, we follow the default settings in randomized SVD and subspace iterations.
We run grid searches for remaining parameters (i.e., $\alpha$ and $T$) and report the best results. More details regarding parameter setup are in Appendix~\ref{sec:add-exp}.
The datasets and our code are available at \url{https://github.com/HKBU-LAGAS/S2CAG}.

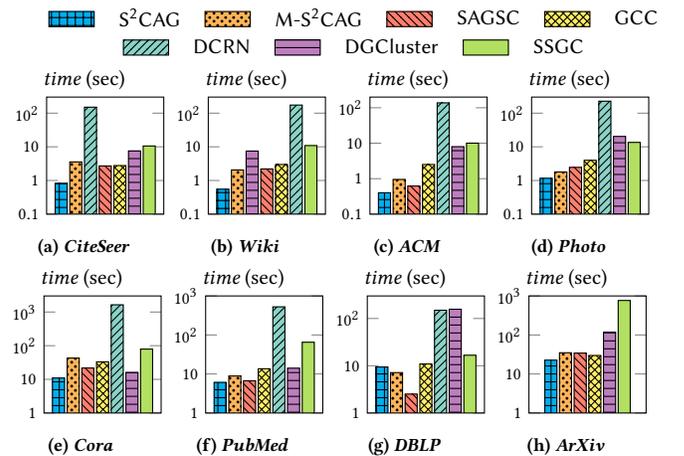
\begin{figure}[!t]
\centering
\begin{small}
\begin{tikzpicture}
   \hspace{3mm}\begin{customlegend}[
        legend entries={\algo,\algoplus ,\textsf{SAGSC},\textsf{GCC}},
        legend columns=4,
        area legend,
        legend style={at={(0.45,1.15)},anchor=north,draw=none,font=\small,column sep=0.25cm}]
        \addlegendimage{preaction={fill, cyan}, pattern={grid}}   
        \addlegendimage{preaction={fill, myorange}, pattern={crosshatch dots}} 
        \addlegendimage{preaction={fill, myred},pattern=north west lines} 
        \addlegendimage{preaction={fill, myyellow},pattern=crosshatch}    
    \end{customlegend}
    \begin{customlegend}[
        legend entries={\textsf{DCRN},\textsf{DGCluster},\textsf{SSGC}},
        legend columns=3,
        area legend,
        legend style={at={(0.45,0.75)},anchor=north,draw=none,font=\small,column sep=0.25cm}]
        \addlegendimage{preaction={fill, mycyan},pattern=north east lines}    
        \addlegendimage{preaction={fill, mypurple},pattern=horizontal lines}    
        \addlegendimage{preaction={fill, mygreen}}    
    \end{customlegend}
\end{tikzpicture}
\\[-\lineskip]
\vspace{-4mm}
\hspace{-2mm}\subfloat[{\em CiteSeer}]{
\begin{tikzpicture}[scale=1]
\begin{axis}[
    height=\columnwidth/2.7,
    width=\columnwidth/2.7,
    xtick=\empty,
    ybar=1.0pt,
    bar width=0.16cm,
    enlarge x limits=true,
    ylabel={\em time} (sec),
    xticklabel=\empty,
    ymin=0.1,
    ymax=300,
    ytick={0.1,1,10,100},
    yticklabels={$0.1$,$1$,$10$,$10^2$},
    ymode=log,
    xticklabel style = {font=\scriptsize},
    yticklabel style = {font=\scriptsize},
    log origin y=infty,
    log basis y={10},
    every axis y label/.style={at={(current axis.north west)},right=5mm,above=0mm},
    legend style={draw=none, at={(1.02,1.02)},anchor=north west,cells={anchor=west},font=\scriptsize},
    legend image code/.code={ \draw [#1] (0cm,-0.1cm) rectangle (0.3cm,0.15cm); },
    ]

\addplot [preaction={fill, cyan}, pattern={grid}] coordinates {(1,0.82) }; 
\addplot [preaction={fill, myorange}, pattern={crosshatch dots}] coordinates {(1,3.56) }; 
\addplot [preaction={fill, mycyan},pattern=north east lines] coordinates {(1,150.09) }; 
\addplot [preaction={fill, myred},pattern=north west lines] coordinates {(1,2.7) }; 
\addplot [preaction={fill, myyellow},pattern=crosshatch] coordinates {(1,2.77) };  
\addplot [preaction={fill, mypurple},pattern=horizontal lines] coordinates {(1,7.54) };
\addplot [preaction={fill, mygreen}] coordinates {(1,10.57) };


\end{axis}
\end{tikzpicture}\hspace{0mm}\label{fig:time-citeseer}%
}%
\subfloat[{\em Wiki}]{
\begin{tikzpicture}[scale=1]
\begin{axis}[
    height=\columnwidth/2.7,
    width=\columnwidth/2.7,
    xtick=\empty,
    ybar=1.0pt,
    bar width=0.16cm,
    enlarge x limits=true,
    ylabel={\em time} (sec),
    xticklabel=\empty,
    ymin=0.1,
    ymax=300,
    ytick={0.1,1,10,100},
    yticklabels={$0.1$,$1$,$10$,$10^2$},
    ymode=log,
    xticklabel style = {font=\scriptsize},
    yticklabel style = {font=\scriptsize},
    log origin y=infty,
    log basis y={10},
    every axis y label/.style={at={(current axis.north west)},right=5mm,above=0mm},
    legend style={draw=none, at={(1.02,1.02)},anchor=north west,cells={anchor=west},font=\scriptsize},
    legend image code/.code={ \draw [#1] (0cm,-0.1cm) rectangle (0.3cm,0.15cm); },
    ]

\addplot [preaction={fill, cyan}, pattern={grid}] coordinates {(1,0.55) }; 
\addplot [preaction={fill, myorange}, pattern={crosshatch dots}] coordinates {(1,2.05) }; 
\addplot [preaction={fill, mypurple},pattern=horizontal lines] coordinates {(1,7.47) }; 
\addplot [preaction={fill, myred},pattern=north west lines] coordinates {(1,2.2) };
\addplot [preaction={fill, myyellow},pattern=crosshatch] coordinates {(1,2.98) }; 
\addplot [preaction={fill, mycyan},pattern=north east lines] coordinates {(1,173.21) };

\addplot [preaction={fill, mygreen}] coordinates {(1,10.87) };


\end{axis}
\end{tikzpicture}\hspace{0mm}\label{fig:time-wiki}%
}%
\subfloat[\em{ACM}]{
\begin{tikzpicture}[scale=1]
\begin{axis}[
    height=\columnwidth/2.7,
    width=\columnwidth/2.7,
    xtick=\empty,
    ybar=1.0pt,
    bar width=0.16cm,
    enlarge x limits=true,
    ylabel={\em time} (sec),
    xticklabel=\empty,
    ymin=0.1,
    ymax=200,
    ytick={0.1,1,10,100},
    yticklabels={$0.1$,$1$,$10$,$10^2$},
    ymode=log,
    xticklabel style = {font=\scriptsize},
    yticklabel style = {font=\scriptsize},
    log origin y=infty,
    log basis y={10},
    every axis y label/.style={at={(current axis.north west)},right=5mm,above=0mm},
    legend style={draw=none, at={(1.02,1.02)},anchor=north west,cells={anchor=west},font=\scriptsize},
    legend image code/.code={ \draw [#1] (0cm,-0.1cm) rectangle (0.3cm,0.15cm); },
    ]

\addplot [preaction={fill, cyan}, pattern={grid}] coordinates {(1,0.4) };
\addplot [preaction={fill, myorange}, pattern={crosshatch dots}] coordinates {(1,0.95) }; 
\addplot [preaction={fill, myred},pattern=north west lines] coordinates {(1,0.62) }; 
\addplot [preaction={fill, myyellow},pattern=crosshatch] coordinates {(1,2.51) };
\addplot [preaction={fill, mycyan},pattern=north east lines] coordinates {(1,137.1) }; 
\addplot [preaction={fill, mypurple},pattern=horizontal lines] coordinates {(1,7.96) }; 
\addplot [preaction={fill, mygreen}] coordinates {(1,10.05) }; 


\end{axis}
\end{tikzpicture}\hspace{0mm}\label{fig:time-acm}%
}%
\subfloat[{\em Photo}]{
\begin{tikzpicture}[scale=1]
\begin{axis}[
    height=\columnwidth/2.7,
    width=\columnwidth/2.7,
    xtick=\empty,
    ybar=1.0pt,
    bar width=0.16cm,
    enlarge x limits=true,
    ylabel={\em time} (sec),
    xticklabel=\empty,
    ymin=0.1,
    ymax=300,
    ytick={0.1,1,10,100},
    yticklabels={$0.1$,$1$,$10$,$10^2$},
    ymode=log,
    xticklabel style = {font=\scriptsize},
    yticklabel style = {font=\scriptsize},
    log origin y=infty,
    log basis y={10},
    every axis y label/.style={at={(current axis.north west)},right=5mm,above=0mm},
    legend style={draw=none, at={(1.02,1.02)},anchor=north west,cells={anchor=west},font=\scriptsize},
    legend image code/.code={ \draw [#1] (0cm,-0.1cm) rectangle (0.3cm,0.15cm); },
    ]

\addplot [preaction={fill, cyan}, pattern={grid}] coordinates {(1,1.16) }; 
\addplot [preaction={fill, myorange}, pattern={crosshatch dots}] coordinates {(1,1.79) }; 
\addplot [preaction={fill, myred},pattern=north west lines] coordinates {(1,2.45) }; 
\addplot [preaction={fill, myyellow},pattern=crosshatch] coordinates {(1,3.98) };  
\addplot [preaction={fill, mycyan},pattern=north east lines] coordinates {(1,226.08) }; 
\addplot [preaction={fill, mypurple},pattern=horizontal lines] coordinates {(1,20.33) };
\addplot [preaction={fill, mygreen}] coordinates {(1,13.52) };


\end{axis}
\end{tikzpicture}\hspace{0mm}\label{fig:time-photos}%
}%
\\[-\lineskip]
\vspace{-3mm}
\hspace{-2mm}\subfloat[{\em Cora}]{
\begin{tikzpicture}[scale=1]
\begin{axis}[
    height=\columnwidth/2.7,
    width=\columnwidth/2.7,
    xtick=\empty,
    ybar=1.0pt,
    bar width=0.16cm,
    enlarge x limits=true,
    ylabel={\em time} (sec),
    xticklabel=\empty,
    ymin=1,
    ymax=3000,
    ytick={1,10,100,1000},
    yticklabels={$1$,$10$,$10^2$,$10^3$},
    ymode=log,
    xticklabel style = {font=\scriptsize},
    yticklabel style = {font=\scriptsize},
    log origin y=infty,
    log basis y={10},
    every axis y label/.style={at={(current axis.north west)},right=5mm,above=0mm},
    legend style={draw=none, at={(1.02,1.02)},anchor=north west,cells={anchor=west},font=\scriptsize},
    legend image code/.code={ \draw [#1] (0cm,-0.1cm) rectangle (0.3cm,0.15cm); },
    ]

\addplot [preaction={fill, cyan}, pattern={grid}] coordinates {(1,10.98) }; 
\addplot [preaction={fill, myorange}, pattern={crosshatch dots}] coordinates {(1,42.94) }; 
\addplot [preaction={fill, myred},pattern=north west lines] coordinates {(1,21.72) }; 
\addplot [preaction={fill, myyellow},pattern=crosshatch] coordinates {(1,33.07) };  
\addplot [preaction={fill, mycyan},pattern=north east lines] coordinates {(1,1632.47) }; 
\addplot [preaction={fill, mypurple},pattern=horizontal lines] coordinates {(1,16.03) };
\addplot [preaction={fill, mygreen}] coordinates {(1,79.2) };


\end{axis}
\end{tikzpicture}\hspace{0mm}\label{fig:time-Cora}%
}%
\subfloat[{\em PubMed}]{
\begin{tikzpicture}[scale=1]
\begin{axis}[
    height=\columnwidth/2.7,
    width=\columnwidth/2.7,
    xtick=\empty,
    ybar=1.0pt,
    bar width=0.16cm,
    enlarge x limits=true,
    ylabel={\em time} (sec),
    xticklabel=\empty,
    ymin=1,
    ymax=1000,
    ymode=log,
    xticklabel style = {font=\scriptsize},
    yticklabel style = {font=\scriptsize},
    ytick={1,10,100,1000},
    yticklabels={$1$,$10$,$10^2$,$10^3$},
    log basis y={10},
    every axis y label/.style={at={(current axis.north west)},right=5mm,above=0mm},
    legend style={draw=none, at={(1.02,1.02)},anchor=north west,cells={anchor=west},font=\scriptsize},
    legend image code/.code={ \draw [#1] (0cm,-0.1cm) rectangle (0.3cm,0.15cm); },
    ]

\addplot [preaction={fill, cyan}, pattern={grid}] coordinates {(1,6.08) }; 
\addplot [preaction={fill, myorange}, pattern={crosshatch dots}] coordinates {(1,8.88) }; 
\addplot [preaction={fill, myred},pattern=north west lines] coordinates {(1,6.69) };
\addplot [preaction={fill, myyellow},pattern=crosshatch] coordinates {(1,13.61) }; 
\addplot [preaction={fill, mycyan},pattern=north east lines] coordinates {(1,524.65) }; 
\addplot [preaction={fill, mypurple},pattern=horizontal lines] coordinates {(1,14.08) };  
\addplot [preaction={fill, mygreen}] coordinates {(1,65.14) };


\end{axis}
\end{tikzpicture}\hspace{0mm}\label{fig:time-pubmed}%
}%
\subfloat[{\em DBLP}]{
\begin{tikzpicture}[scale=1]
\begin{axis}[
    height=\columnwidth/2.7,
    width=\columnwidth/2.7,
    xtick=\empty,
    ybar=1.0pt,
    bar width=0.16cm,
    enlarge x limits=true,
    ylabel={\em time} (sec),
    xticklabel=\empty,
    ymin=1,
    ymax=300,
    ytick={1,10,100},
    yticklabels={$1$,$10$,$10^2$},
    ymode=log,
    xticklabel style = {font=\scriptsize},
    yticklabel style = {font=\scriptsize},
    log origin y=infty,
    log basis y={10},
    every axis y label/.style={at={(current axis.north west)},right=5mm,above=0mm},
    legend style={draw=none, at={(1.02,1.02)},anchor=north west,cells={anchor=west},font=\scriptsize},
    legend image code/.code={ \draw [#1] (0cm,-0.1cm) rectangle (0.3cm,0.15cm); },
    ]

\addplot [preaction={fill, cyan}, pattern={grid}] coordinates {(1,9.42) }; 
\addplot [preaction={fill, myorange}, pattern={crosshatch dots}] coordinates {(1,7.15) }; 
\addplot [preaction={fill, myred},pattern=north west lines] coordinates {(1,2.51) }; 
\addplot [preaction={fill, myyellow},pattern=crosshatch] coordinates {(1,10.92) };  
\addplot [preaction={fill, mycyan},pattern=north east lines] coordinates {(1,149.51) }; 
\addplot [preaction={fill, mypurple},pattern=horizontal lines] coordinates {(1,157.14) };
\addplot [preaction={fill, mygreen}] coordinates {(1,16.7) };


\end{axis}
\end{tikzpicture}\hspace{0mm}\label{fig:time-dblp}%
}%
\subfloat[{\em ArXiv}]{
\begin{tikzpicture}[scale=1]
\begin{axis}[
    height=\columnwidth/2.7,
    width=\columnwidth/2.7,
    xtick=\empty,
    ybar=1.0pt,
    bar width=0.16cm,
    enlarge x limits=true,
    ylabel={\em time} (sec),
    xticklabel=\empty,
    ymin=1,
    ymax=1000,
    ytick={1,10,100,1000},
    yticklabels={$1$,$10$,$10^2$,$10^3$},
    ymode=log,
    xticklabel style = {font=\scriptsize},
    yticklabel style = {font=\scriptsize},
    log origin y=infty,
    log basis y={10},
    every axis y label/.style={at={(current axis.north west)},right=5mm,above=0mm},
    legend style={draw=none, at={(1.02,1.02)},anchor=north west,cells={anchor=west},font=\scriptsize},
    legend image code/.code={ \draw [#1] (0cm,-0.1cm) rectangle (0.3cm,0.15cm); },
    ]

\addplot [preaction={fill, cyan}, pattern={grid}] coordinates {(1,22.63) }; 
\addplot [preaction={fill, myorange}, pattern={crosshatch dots}] coordinates {(1,34.96) }; 
\addplot [preaction={fill, myred},pattern=north west lines] coordinates {(1,34.31) }; 
\addplot [preaction={fill, myyellow},pattern=crosshatch] coordinates {(1,29.8) };  
\addplot [preaction={fill, mypurple},pattern=horizontal lines] coordinates {(1,118.37) };
\addplot [preaction={fill, mygreen}] coordinates {(1,765.37) };


\end{axis}
\end{tikzpicture}\hspace{0mm}\label{fig:time-arxiv}%
}%
\vspace{-4mm}
\end{small}
\caption{Clustering efficiency performance.} \label{fig:time}
\vspace{-3ex}
\end{figure}

\subsection{Performance Evaluation}
\subsubsection{\bf Clustering Quality}
This set of experiments reports the ACC, NMI, and ARI scores achieved by \algo, \algoplus, and all baselines over 8 datasets. We conduct 5 trials and report the averaged values and standard deviation over the trials. A method is omitted if it fails to report the results within one day or incurs out-of-memory errors.

Tables~\ref{tbl:node-clustering-small} and ~\ref{tbl:node-clustering-large} show the ACC, NMI, ARI scores, and the average performance rankings. The best and second-best results are highlighted in blue and darker shades indicate better clustering. The best baselines are underlined. From the last columns of Tables~\ref{tbl:node-clustering-small} and ~\ref{tbl:node-clustering-large}, we can see that \algoplus and \algo are ranked the highest and second highest in terms of overall clustering quality among all evaluated methods, respectively, whereas the best performer \textsf{SAGSC} \cite{fettal2023scalable} in baselines is another SCAG approach.
This observation validates the effectiveness of subspace clustering for attributed graph clustering tasks.

Specifically, on small and medium-sized attributed graphs, \algo and \algoplus outperform all competing methods, with conspicuous improvements pertinent to almost all metrics. For instance, our proposed methods improve the best baselines by significant margins of $8\%$, $4.9\%$, $4.3\%$ in ACC, NMI, and ARI on {\em Wiki}, and $4.4\%$, $2.1\%$, $8.2\%$ on {\em PubMed}, respectively. This superiority is still pronounced on datasets with millions of edges, i.e., {\em DBLP} and {\em ArXiv}, where we can observe that \algoplus is able to give a performance gain of $0.4\%$, $2.2\%$ for ACC and $1.0\%$, $2.3\%$ for ARI, respectively.
Moreover, on all datasets except a small one {\em Wiki}, it can be observed that \algoplus yields comparable and often superior performance to \algo.
Overall, \algoplus improves \algo in generalization and robustness by maximizing modularity.

\subsubsection{\bf Efficiency}

Fig.~\ref{fig:time} depicts the running times required by \algo, \algoplus, and other competitive baselines (ranked top 7 in Tables~\ref{tbl:node-clustering-small} and ~\ref{tbl:node-clustering-large}) for clustering on all datasets. 
The $y$-axis represents the running time (seconds) in log scale.
The reported runtime values exclude the costs for input (loading datasets) and output (saving clustering results).

From Fig.~\ref{fig:time-citeseer} and ~\ref{fig:time-wiki}, we can observe that \algo is able to gain $183\times$ and $13.6\times$ speedup on {\em CiteSeer} and {\em Wiki} when compared to the best baselines \textsf{DCRN} and \textsf{DGCluster}, respectively. On the rest of the datasets except {\em DBLP}, \algo and \algoplus achieve comparable efficiency to the state of the art \textsf{SAGSC} and are among the fastest methods. Over the {\em DBLP} graph with 2.5 million edges, our solutions take at least 7.2 seconds to terminate and \textsf{SAGSC} consumes 2.5 seconds. But recall that in Table~\ref{tbl:node-clustering-large}, \algo and \algoplus outperform \textsf{SAGSC} by a considerable gain of $0.4\%$, $0.8\%$, $1.0\%$ in ACC, NMI, and ARI, respectively.

In summary, \algo and \algoplus consistently deliver superior results for clustering on various attributed graphs while offering high empirical efficiency. The empirical observation corroborates the efficacy of our novel objective function in Section~\ref{sec:SCAG} and algorithmic designs developed in Sections~\ref{sec:algo} and Section~\ref{sec:algo-plus}.

\subsection{Parameter Study}
This set of experiments investigates the effects of parameters $\alpha$, $T$, and $\gamma$ in \algo and \algoplus. 
We run \algo and \algoplus on the three largest datasets {\em PubMed}, {\em DBLP}, and {\em ArXiv}, respectively, by varying each parameter while fixing others as in Section~\ref{sec:exp-setup}.

\begin{figure}[!t]
\centering
\begin{small}
\begin{tikzpicture}
    \begin{customlegend}
    [legend columns=2,
        legend entries={\algo, \algoplus},
        legend style={at={(0.45,1.35)},anchor=north,draw=none,font=\footnotesize,column sep=0.2cm}]
    \addlegendimage{line width=0.2mm,mark size=4pt,mark=triangle,color=blue}
    \addlegendimage{line width=0.2mm,mark size=4pt,mark=diamond,color=orange}
    \end{customlegend}
\end{tikzpicture}
\\[-\lineskip]
\vspace{-4mm}
\subfloat[\em PubMed]{
\begin{tikzpicture}[scale=1,every mark/.append style={mark size=2pt}]
    \begin{axis}[
        height=\columnwidth/2.8,
        width=\columnwidth/2.3,
        ylabel={\it ACC},
        xmin=0.5, xmax=11.5,
        ymin=0.5, ymax=0.8,
        xtick={1,3,5,7,9,11},
        ytick={0.5,0.6,0.7,0.8},
        xticklabel style = {font=\scriptsize},
        yticklabel style = {font=\footnotesize},
        xticklabels={0.2,0.6,1.0,1.4,1.8,2.2},
        yticklabels={0.5,0.6,0.7,0.8},
        every axis y label/.style={font=\footnotesize,at={(current axis.north west)},right=2mm,above=0mm},
        legend style={fill=none,font=\small,at={(0.02,0.99)},anchor=north west,draw=none},
    ]
    \addplot[line width=0.3mm, mark=triangle,color=blue]  
        plot coordinates {
(1,	0.5507	)
(3,	0.5572	)
(5,	0.6892	)
(7,	0.751	)
(9,	0.7528	)
(11,	0.7512	)
};

    \addplot[line width=0.3mm, mark=diamond,color=orange]  
        plot coordinates {
(1,	0.598	)
(3,	0.6022	)
(5,	0.6892	)
(7,	0.7545	)
(9,	0.755	)
(11,	0.7549	)

    };

    \end{axis}
\end{tikzpicture}\hspace{0mm}\label{fig:alpha-pubmed}%
}
\subfloat[\em DBLP]{
\begin{tikzpicture}[scale=1,every mark/.append style={mark size=2pt}]
    \begin{axis}[
        height=\columnwidth/2.8,
        width=\columnwidth/2.3,
        ylabel={\it ACC},
        xmin=0.5, xmax=11.5,
        ymin=0.6, ymax=1.0,
        xtick={1,3,5,7,9,11},
        ytick={0.5,0.6,0.7,0.8,0.9,1.0},
        xticklabel style = {font=\scriptsize},
        yticklabel style = {font=\footnotesize},
        xticklabels={0.1,0.3,0.5,0.7,0.9,1.2},
        yticklabels={0.5,0.6,0.7,0.8,0.9,1.0},
        every axis y label/.style={font=\footnotesize,at={(current axis.north west)},right=2mm,above=0mm},
        legend style={fill=none,font=\small,at={(0.02,0.99)},anchor=north west,draw=none},
    ]
    \addplot[line width=0.3mm, mark=triangle,color=blue]  
        plot coordinates {
(1,	0.6357	)
(3,	0.8349	)
(5,	0.8807	)
(7,	0.9159	)
(9,	0.9354	)
(11,	0.9334	)

    };

   \addplot[line width=0.3mm, mark=diamond,color=orange]  
        plot coordinates {
(1,	0.8038 )
(3,	0.8376	)
(5,	0.8802	)
(7,	0.913	)
(9,	0.9352	)
(12,	0.9349	)

};

    \end{axis}
\end{tikzpicture}\hspace{0mm}\label{fig:alpha-dblp}%
}
\subfloat[\em ArXiv]{
\begin{tikzpicture}[scale=1,every mark/.append style={mark size=2pt}]
    \begin{axis}[
        height=\columnwidth/2.8,
        width=\columnwidth/2.3,
        ylabel={\it ACC},
        xmin=0.5, xmax=11.5,
        ymin=0.2, ymax=0.6,
        xtick={1,3,5,7,9,11},
        ytick={0.1,0.2,0.3,0.4,0.5,0.6},
        xticklabel style = {font=\scriptsize},
        yticklabel style = {font=\footnotesize},
        xticklabels={0.5,1.0,1.5,2.0,2.5,3.0},
        yticklabels={0.1,0.2,0.3,0.4,0.5,0.6},
        every axis y label/.style={font=\footnotesize,at={(current axis.north west)},right=2mm,above=0mm},
        legend style={fill=none,font=\small,at={(0.02,0.99)},anchor=north west,draw=none},
    ]
    \addplot[line width=0.3mm, mark=triangle,color=blue]  
        plot coordinates {
(1,	0.3105	)
(3,	0.4429	)
(5,	0.4686	)
(7,	0.461	)
(9,	0.462	)
(11,	0.460	)
    };

    \addplot[line width=0.3mm, mark=diamond,color=orange]  
        plot coordinates {
(1,	0.3208	)
(3,	0.4263	)
(5,	0.488	)
(7,	0.4939	)
(9,	0.4997	)
(11,	0.4927	)

    };

    \end{axis}
\end{tikzpicture}\hspace{0mm}\label{fig:alpha=arxiv}%
}

\end{small}
 \vspace{-4mm}
\caption{Clustering accuracy when varying $\alpha$} \label{fig:alpha}
\vspace{-5ex}
\end{figure}
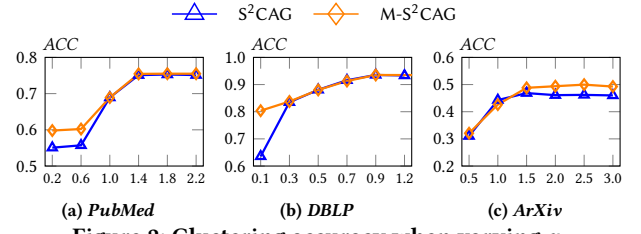

\begin{figure}[!t]
\centering
\begin{small}
\subfloat[\em PubMed]{
\begin{tikzpicture}[scale=1,every mark/.append style={mark size=2pt}]
    \begin{axis}[
        height=\columnwidth/2.8,
        width=\columnwidth/2.35,
        ylabel={\it ACC},
        xmin=0.5, xmax=11.5,
        ymin=0.65, ymax=0.8,
        xtick={1,3,5,7,9,11},
        ytick={0.65,0.7,0.75,0.8},
        xticklabel style = {font=\scriptsize},
        yticklabel style = {font=\footnotesize},
        xticklabels={50,100,125,150,175,200},
        yticklabels={0.65,0.7,0.75,0.8},
        every axis y label/.style={font=\footnotesize,at={(current axis.north west)},right=2mm,above=0mm},
        legend style={fill=none,font=\small,at={(0.02,0.99)},anchor=north west,draw=none},
    ]
    \addplot[line width=0.3mm, mark=triangle,color=blue]  
        plot coordinates {
(1,	0.6808	)
(3,	0.6942	)
(5,	0.7489	)
(7,	0.7495	)
(9,	0.7528	)
(11,	0.7438	)

    };
    \addplot[line width=0.3mm, mark=diamond,color=orange]  
        plot coordinates {
(1,	0.6805	)
(3,	0.7423	)
(5,	0.7489	)
(7,	0.7504	)
(9,	0.755	)
(11,	0.7538	)
    };

    \end{axis}
\end{tikzpicture}\hspace{0mm}\label{fig:iter-pubmed}%
}
\subfloat[\em DBLP]{
\begin{tikzpicture}[scale=1,every mark/.append style={mark size=2pt}]
    \begin{axis}[
        height=\columnwidth/2.8,
        width=\columnwidth/2.35,
        ylabel={\it ACC},
        xmin=0.5, xmax=11.5,
        ymin=0.85, ymax=0.95,
        xtick={1,3,5,7,9,11},
        ytick={0.8,0.85,0.9,0.95,1.0},
        xticklabel style = {font=\scriptsize},
        yticklabel style = {font=\footnotesize},
        xticklabels={1,5,10,20,30,40},
        yticklabels={0.8,0.85,0.9,0.95,1.0},
        every axis y label/.style={font=\footnotesize,at={(current axis.north west)},right=2mm,above=0mm},
        legend style={fill=none,font=\small,at={(0.02,0.99)},anchor=north west,draw=none},
    ]
    \addplot[line width=0.3mm, mark=triangle,color=blue]  
        plot coordinates {
(1,	0.8543	)
(3,	0.9295	)
(5,	0.9354	)
(7,	0.9352	)
(9,	0.9349	)
(11,	0.9344	)
    };

    \addplot[line width=0.3mm, mark=diamond,color=orange]  
        plot coordinates {
(1,	0.8627	)
(3,	0.9305	)
(5,	0.9352	)
(7,	0.9352	)
(9,	0.9347	)
(11,	0.9349	)
    };

    \end{axis}
\end{tikzpicture}\hspace{0mm}\label{fig:iter-dblp}%
}
\subfloat[\em ArXiv]{
\begin{tikzpicture}[scale=1,every mark/.append style={mark size=2pt}]
    \begin{axis}[
        height=\columnwidth/2.8,
        width=\columnwidth/2.35,
        ylabel={\it ACC},
        xmin=0.5, xmax=11.5,
        ymin=0.35, ymax=0.5,
        xtick={1,3,5,7,9,11},
        ytick={0.3,0.35,0.4,0.45,0.5},
        xticklabel style = {font=\scriptsize},
        yticklabel style = {font=\footnotesize},
        xticklabels={5,10,20,25,30,35},
        yticklabels={0.3,0.35,0.4,0.45,0.5},
        every axis y label/.style={font=\footnotesize,at={(current axis.north west)},right=2mm,above=0mm},
        legend style={fill=none,font=\small,at={(0.02,0.99)},anchor=north west,draw=none},
    ]
    \addplot[line width=0.3mm, mark=triangle,color=blue]  
        plot coordinates {
(1,	0.365	)
(3,	0.4166	)
(5,	0.4339	)
(7,	0.4607	)
(9,	0.4686	)
(11,	0.4535	)
    };
    \addplot[line width=0.3mm, mark=diamond,color=orange]  
        plot coordinates {
(1,	0.3891	)
(3,	0.4142	)
(5,	0.4433	)
(7,	0.4726	)
(9,	0.4997	)
(11,	0.465	)
    };

    \end{axis}
\end{tikzpicture}\hspace{0mm}\label{fig:iter-arxiv}%
}%

\end{small}
 \vspace{-4mm}
\caption{Clustering accuracy when varying $T$.} \label{fig:iter}
\vspace{-5ex}
\end{figure}
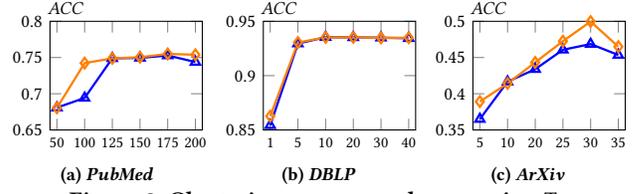

\stitle{Varying $\alpha$} 
Fig. \ref{fig:alpha} illustrates the ACC scores obtained by \algo and \algoplus on {\em PubMed}, {\em DBLP}, and {\em ArXiv} when $\alpha$ is varied from $0.2$ to $2.2$, $0.1$ to $1.2$, and $0.5$ to $3$, respectively.
It can be observed that on all tested datasets, both \algo and \algoplus see a drastic uptick in ACC and reach a plateau afterward when increasing $\alpha$. Specifically, when $\alpha$ is beyond $1.4$, $0.9$, and $2.0$, the ACC scores remain almost invariant on {\em PubMed}, {\em DBLP}, and {\em ArXiv}, respectively. These observations reveal that large $\alpha$ (especially over $1.0$) is conducive for clustering, as it can amplify the feature patterns of far-reaching neighbors in Eq.~\eqref{eq:NSR}, which are restrained in standard vertex representations calculated in Eq.~\eqref{eq:smooth-x} by limiting $\alpha$ within $(0,1)$.

\stitle{Varying $T$} Fig.~\ref{fig:iter} plots the ACC scores of \algo and \algoplus when varying order $T$ from $50$ to $200$, $1$ to $40$, and $5$ to $35$, on {\em PubMed}, {\em DBLP}, and {\em ArXiv}, respectively. Observe that \algo and \algoplus experience a rapid improvement in ACC with $T$ increasing in the beginning. Subsequently, the clustering quality of both of them remains relatively stable on {\em PubMed} and {\em DBLP}, but witnesses a sharp performance decline on {\em ArXiv} when $T$ exceeds $30$. 
This phenomenon is attributed to the {\em over-smoothing}~\cite{li2018deeper} and {\em over-squashing}~\cite{alonbottleneck} issues caused by large orders $T$.
However, on {\em PubMed}, our methods require an order $T$ greater than 100 to attain satisfactory performance since vertices inside it are poorly connected.

\begin{figure}[!t]
\centering
\begin{small}
\subfloat[\em PubMed]{
\begin{tikzpicture}[scale=1,every mark/.append style={mark size=2pt}]
    \begin{axis}[
        height=\columnwidth/2.8,
        width=\columnwidth/2.3,
        ylabel={\it ACC},
        xmin=0.5, xmax=11.5,
        ymin=0.5, ymax=0.8,
        xtick={1,3,5,7,9,11},
        ytick={0.5,0.6,0.7,0.8},
        xticklabel style = {font=\scriptsize},
        yticklabel style = {font=\footnotesize},
        xticklabels={0.7,0.8,0.9,1.0,1.1,1.2},
        yticklabels={0.5,0.6,0.7,0.8},
        every axis y label/.style={font=\footnotesize,at={(current axis.north west)},right=2mm,above=0mm},
        legend style={fill=none,font=\small,at={(0.02,0.99)},anchor=north west,draw=none},
    ]

 \addplot[line width=0.3mm, mark=diamond,color=orange]  
        plot coordinates {
(1,	0.592	)
(3,	0.592	)
(5,	0.592	)
(7,	0.755	)
(9,	0.592	)
(11,	0.592	)
    };

    \end{axis}
\end{tikzpicture}\hspace{-1mm}\label{fig:gamma-pubmed}
}
\subfloat[\em DBLP]{
\begin{tikzpicture}[scale=1,every mark/.append style={mark size=2pt}]
    \begin{axis}[
        height=\columnwidth/2.8,
        width=\columnwidth/2.3,
        ylabel={\it ACC},
        xmin=0.5, xmax=11.5,
        ymin=0.7, ymax=1.0,
        xtick={1,3,5,7,9,11},
        ytick={0.7,0.8,0.9,1.0},
        xticklabel style = {font=\scriptsize},
        yticklabel style = {font=\footnotesize},
        xticklabels={0.7,0.8,0.9,1.0,1.1,1.2},
        yticklabels={0.7,0.8,0.9,1.0},
        every axis y label/.style={font=\footnotesize,at={(current axis.north west)},right=2mm,above=0mm},
        legend style={fill=none,font=\small,at={(0.02,0.99)},anchor=north west,draw=none},
    ]
    \addplot[line width=0.3mm, mark=diamond,color=orange]  
        plot coordinates {
(1,	0.7826	)
(3,	0.7826	)
(5,	0.7069	)
(7,	0.9352	)
(9,	0.7069	)
(11,	0.7826	)
    };

    \end{axis}
\end{tikzpicture}\hspace{-1mm}\label{fig:gamma-dblp}
}
\subfloat[\em ArXiv]{
\begin{tikzpicture}[scale=1,every mark/.append style={mark size=2pt}]
    \begin{axis}[
        height=\columnwidth/2.8,
        width=\columnwidth/2.3,
        ylabel={\it ACC},
        xmin=0.5, xmax=11.5,
        ymin=0.47, ymax=0.50,
        xtick={1,3,5,7,9,11},
        ytick={0.47,0.48,0.49,0.50},
        xticklabel style = {font=\scriptsize},
        yticklabel style = {font=\footnotesize},
        xticklabels={0.7,0.8,0.9,1.0,1.1,1.2},
        yticklabels={0.47,0.48,0.49,0.50},
        every axis y label/.style={font=\footnotesize,at={(current axis.north west)},right=2mm,above=0mm},
        legend style={fill=none,font=\small,at={(0.02,0.99)},anchor=north west,draw=none},
    ]

 \addplot[line width=0.3mm, mark=diamond,color=orange]  
        plot coordinates {
(1,	0.4826	)
(3,	0.4826	)
(5,	0.4825	)
(7,	0.4997	)
(9,	0.4825	)
(11,	0.4826	)
    };

    \end{axis}
\end{tikzpicture}\hspace{0mm}\label{fig:gamma-arxiv}
}
\end{small}
 \vspace{-4mm}
\caption{Varying $\gamma$ in \algoplus \label{fig:gamma} }
\vspace{-3ex}
\end{figure}
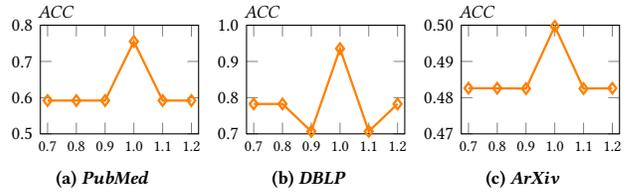

\stitle{Varying $\gamma$} As shown in Fig.~\ref{fig:gamma}, \algoplus achieves the best ACC when $\gamma$ is around $1.0$, which is consistent with the original definition of modularity discussed in Section~\ref{sec:modularity}. Moreover, it can be observed from Fig.~\ref{fig:gamma} that \algoplus is highly sensitive to $\gamma$, whose performance is considerably inferior when $\gamma$ is only slightly smaller or greater than $1.0$. The reason is that the affinity values $\hat{\ZM}_i\cdot\hat{\ZM}_j\ \forall{v_i,v_j}$ in Eq.~\eqref{eq:modularity-W} are small and subtly different due to the normalization. Hence, if $\gamma$ is large, these affinity values will be masked by $\textstyle \gamma\cdot \frac{\wvec_i\cdot \wvec_j}{{\wvec^{\top}\boldsymbol{1}}}$, and they remain indistinguishable when $\gamma$ is small.

\section{Conclusion}
In this paper, we present two effective and scalable solutions, \algo and \algoplus, for SCAG. Under the hood, our proposed methods include (i) a new optimization objective built on an optimized representation model and non-trivial constraints, (ii) fast and theoretically-grounded optimization solvers, and (iii) careful theoretical analyses investigating the rationale underlying \algo and \algoplus.
Our thorough evaluation results manifest the efficacy of our techniques in addressing the limitations of existing works for vertex clustering over attributed graph datasets of varied volumes.

\begin{acks}
Renchi Yang is supported by the NSFC Young Scientists Fund (No. 62302414) and Hong Kong RGC ECS grant (No. 22202623). Xiangyu Ke is supported by the Ningbo Yongjiang Talent Introduction Programme (2022A-237-G) and Zhejiang Province's ``Lingyan'' R\&D Project under Grant No. 2024C01259.
\end{acks}

\balance
\bibliographystyle{ACM-Reference-Format}
\bibliography{main}

\appendix
\section{Algorithmic Details}\label{sec:add-detail}

\subsection{\itrgf}\label{sec:power-iter}
Algo.~\ref{alg:power} displays the pseudo-code of \itrgf for computing $\ZM$ defined in Eq.~\eqref{eq:NSR} in an iterative manner. Specifically, after taking as inputs matrices $\hat{\PM}$, $\hat{\XM}$, an integer $T$, and decay factor $\alpha$, Algo.~\ref{alg:power} initializes $\ZM$ as $\frac{1-\alpha}{1-\alpha^{T+1}}\cdot \hat{\XM}$. Later, \algo starts $T$ rounds of matrix multiplications and each round updates $\ZM$ by
\begin{equation}\label{eq:update-Z}
\textstyle \ZM \gets \alpha \cdot \hat{\PM} \ZM + \frac{1-\alpha}{1-\alpha^{T+1}}\cdot \hat{\XM},
\end{equation}
and returns the $\ZM$ at the end of $T$-th iteration as the output.

\begin{algorithm}[!h]
\caption{\itrgf}\label{alg:power}
\KwIn{Matrices $\hat{\PM}$, $\hat{\XM}$, order $T$, decay factor $\alpha$}
\KwOut{NSR $\ZM$}
$\ZM \gets \frac{1-\alpha}{1-\alpha^{T+1}}\cdot \hat{\XM}$;\\
\For{$t\gets 1$ \KwTo $T$ }{
 update $\ZM$ according to Eq.~\eqref{eq:update-Z};\\
}
\end{algorithm}

Since $\hat{\PM}$ is a sparse matrix containing $m$ non-zero entries, the sparse matrix multiplication $\hat{\PM} \ZM$ can be done in $O(dm)$ time. The total cost is hence $O(Tdm)$ for $T$ iterations.

\subsection{Complexity Analyses}\label{sec:algo-add}
\stitle{\algo}
First, we consider the naive method at Lines 2-3 of Algo.~\ref{alg:sscag}. According to Section~\ref{sec:power-iter}, the construction of NSR $\ZM$ at Line 2 takes $O(Tdm)$. As for Line 3, our implementation of the randomized SVD algorithm~\cite{Halko2009FindingSW} involves $2\cdot (\tau+1)\cdot (k+o)dn + 3\cdot(k+o)^2n$ matrix operations. In sum, the empirical cost can be estimated by
\begin{equation*}
f_{\text{na\"{\i}ve}}(\G,k,\tau,T)=Tdm+2(\tau+1)\cdot (k+o)dn + 3(k+o)^2n.
\end{equation*}
The asymptotic complexity can be simplified as $O(Tdm + k\tau dn)$ since the oversampling parameter $o$ can be regarded as a constant.

The major cost of the integrated method lies at Lines 6-13 of Algo.~\ref{alg:sscag}. As per Section~\ref{sec:power-iter}, for each of the $\tau$ iterations, both Lines 7 and 8 incur $T(k+o)m$ operations when executing \textsf{PowerMethod}, while Lines 7 and 9 need $(k+o)dn$ operations for computing $\hat{\XM}\RM$ and $\hat{\XM}^{\top}\HM$. Similarly, we can analyze that the cost of Lines 10 and 12 is $2T(k+o)m + 2(k+o)dn$.
To sum up, Lines 6-9, 10, and 12 together involve $2(\tau +1)\cdot (k+o)\cdot(dn+Tm)$ operations in total. 
Since the QR decomposition and SVD are applied to $n\times (k+o)$ and $(k+o)\times n$ matrices, respectively, both their runtime can be bounded by $O((k+o)^2n)$. The matrix multiplication in Eq.~\eqref{eq:comp-QY} also requires $O((k+o)^2n)$ operations. Overall, the estimated computational cost of the integrated method can be calculated by
\begin{equation}\label{eq:c1}
f_{\text{integr}}(\G,k,\tau,T)=2(\tau+1)\cdot (k+o) \cdot(dn+Tkm)+3(k+o)^2n,
\end{equation}
which leads to a theoretical time complexity of $O(\tau (kdn+Tm))$ when regarding $o$ as a constant.
By adaptively selecting these two approaches to run, the runtime complexity of \algo is guaranteed to be bounded by $O\left(k\tau dn +\min\{\tau Tm, Tdm\}\right)$.

\stitle{\algoplus} Line 1 invokes Algo.~\ref{alg:power} with $T$ iterations, and hence, takes $O(Tdm)$ time. The computation of $\hat{\ZM}$ at Line 2 can be done in $O(dn)$ time since we can first calculate vector $\zvec = \sum_{v_j\in \V}{\ZM_j}$, followed by a normalization operation $\frac{\ZM_i}{\ZM_i\cdot \zvec^{\top}}$ for each $v_i\in \V$. Similarly, we can compute $\wvec$ by setting $\wvec_i = \hat{\ZM}_i\cdot \hat{\boldsymbol{z}}^{\top}$ for each vertex $v_i\in\V$, where $\hat{\boldsymbol{z}}$ is a sum of all row vectors in $\hat{\ZM}$, i.e., $\hat{\boldsymbol{z}}=\sum_{v_j\in \V}{\hat{\ZM}_j}$. The cost is also $O(dn)$. Both Lines 4 and 7 apply a QR decomposition of an $n\times k$ matrix, requiring $O(k^2n)$ time. Note that Eq.~\eqref{eq:update-H-Q} can be computed in $O(kdn)$ via re-ordered matrix multiplications. Considering all the $\tau$ iterations, the total cost for updating $\HM$ and $\QM$ is then $O(k\tau dn)$. Overall, the runtime complexity of \algoplus is bounded by $O(Tdm+k\tau dn)$.

\begin{table}[!h]
\centering
\renewcommand{\arraystretch}{0.9}
\begin{small}
\caption{Statistics for $\beta_\ell\ \forall{v_\ell\in \V}$.}\label{tbl:X}
\vspace{-4mm}
\begin{tabular}{c|c|c|c|c|c}
	\hline
	 & \multicolumn{1}{c|}{\bf {\em ACM} } & \multicolumn{1}{c|}{\bf {\em Wiki} } & \multicolumn{1}{c|}{\bf {\em CiteSeer} } & \multicolumn{1}{c|}{\bf {\em Photo}}  & \multicolumn{1}{c}{\bf {\em DBLP} } \\
	\hline
   Mean & 0.996 & 0.986 & 0.99 & 0.967 & 0.951 \\
   Variance & 0.004 & 0.013 & 0.009 & 0.031 & 0.044 \\
   \hline
\end{tabular}%
\end{small}
\vspace{-3ex}
\end{table}

\begin{table}[!h]
\centering
\renewcommand{\arraystretch}{0.9}
\begin{small}
\caption{Statistics for $\sum_{v_j\in \V}\ZM_i\cdot\ZM_j^{\top}\ \forall{v_i\in \V}$.}\label{tbl:Z}
\vspace{-4mm}
\begin{tabular}{c|c|c|c|c|c}
	\hline
	 & \multicolumn{1}{c|}{\bf {\em ACM} } & \multicolumn{1}{c|}{\bf {\em Wiki} } & \multicolumn{1}{c|}{\bf {\em CiteSeer} } & \multicolumn{1}{c|}{\bf {\em Photo}}  & \multicolumn{1}{c}{\bf {\em DBLP} } \\
	\hline
   Mean & 0.64 & 1.88 & 0.63 & 1.76 & 0.788 \\
   Variance & 9.2e-4 & 0.04 & 2.62e-3 & 0.019 & 3.11e-3 \\
   \hline
\end{tabular}%
\end{small}
\vspace{-3ex}
\end{table}

\subsection{Stochasticity Analysis of $\ZM\ZM^{\top}$}\label{sec:ZZ-stochastic}
For ease of exposition, for any vertex $v_\ell \in \V$, we first define 
\begin{small}
\begin{equation*}
\begin{aligned}
\beta_\ell=\sum_{v_h\in \V}\hat{\XM}_\ell\cdot\hat{\XM}_h^{\top}\ \text{and}\ \pi_\ell=\sum_{v_h\in \V}\sum_{t=0}^{T}{\frac{\alpha^t}{\sum_{l=0}^{T}{\alpha^l}}\hat{\PM}^{t}_{h,\ell}}.
\end{aligned}
\end{equation*}
\end{small}
\begin{lemma}\label{lem:zz-range}
$\forall{v_i}\in\V$, $\underset{v_\ell\in \V}{\min}{\beta_\ell \cdot \pi_\ell} \le \underset{v_j\in \V}{\sum}\ZM_i{\ZM_j}^{\top} \le \underset{v_\ell\in \V}{\max}{\beta_\ell \cdot \pi_\ell}$.
\end{lemma}
Lemma~\ref{lem:zz-range} provides upper and lower bounds for the sum of entries in each row of $\ZM\ZM^{\top}$, which are dependent on the maximums and minimums of $\beta_\ell$ and $\pi_\ell$. In what follows, we show that the variables in $\{\beta_\ell|v_\ell\in \V\}$ and $\{\pi_\ell|v_\ell\in \V\}$ are dispersed in a narrow value range with low variances, making $\sum_{v_j\in \V}\ZM_i{\ZM_j}^{\top}\ \forall{v_i\in \V}$ nearly identical.

\begin{lemma}\label{lem:XX-range}
$\sum_{v_\ell\in \V}{\beta_\ell} = n$.
\end{lemma}
Theoretically, the above lemma states that the average value of $\beta_\ell\ \forall{v_\ell\in \V}$ is $1$, which accords with its empirical means reported in Table~\ref{tbl:X}. Additionally, the variances of $\beta_\ell\ \forall{v_\ell\in \V}$ on the five datasets are almost negligible, implying that $\min_{v_\ell}{\beta_\ell}\approx \max_{v_\ell}{\beta_\ell}$.

\begin{lemma}\label{lem:pisum}
For any vertex $v_\ell$, the following inequality
\begin{small}
\begin{equation*}
\frac{\hat{d}(v_\ell)}{\max_{v_h\in \N_{v_\ell} \cap \{v_\ell\} }{\hat{d}(v_h)}} \le \pi_\ell \le \frac{\hat{d}(v_\ell)}{\min_{v_h\in \N_{v_\ell} \cap \{v_\ell\} }{\hat{d}(v_h)}},
\end{equation*}
\end{small}
holds, where $\hat{d}(v_h)=\sum_{v_j\in \V}{\NAM_{h,j}}$ and $\hat{d}(v_\ell)=\sum_{v_j\in \V}{\NAM_{\ell,j}}$.
\end{lemma}
Our task thus turns to bound $\pi_\ell\ \forall{v_\ell\in \V}$, which is done in Lemma~\ref{lem:pisum}. Since $\NAM$ is normalized, the values in $\{\hat{d}(v_\ell)| v_\ell\in \V\}$ has a little variation, namely $\min_{v_\ell}{\pi_\ell}$ and $\max_{v_\ell}{\pi_\ell}$ are close.
As a consequence, the row sums $\sum_{v_j\in \V}\ZM_i\cdot\ZM_j^{\top}\ \forall{v_i\in \V}$ are closely aligned, which can be corroborated by our empirical observations on each dataset from Table~\ref{tbl:Z} and indicates that $\ZM\ZM^{\top}$ is nearly a {\em scaled} stochastic matrix.

\section{Additional Experiments}\label{sec:add-exp}

Table~\ref{tbl:param} lists the detailed settings of parameters used in \algo and \algoplus, including decay factor $\alpha$, order $T$, the number of iterations $\tau$, and weight $\gamma$.

\begin{table}[!h]
\centering
\renewcommand{\arraystretch}{0.9}
\begin{small}
\caption{Parameter Settings in \algo/\algoplus.}\label{tbl:param}
\vspace{-3mm}
\resizebox{\columnwidth}{!}{%
\begin{tabular}{c|c|c|c|c|c|c|c|c}
	\hline
	 & \multicolumn{1}{c|}{\bf {\em ACM} } & \multicolumn{1}{c|}{\bf {\em Wiki} } & \multicolumn{1}{c|}{\bf {\em CiteSeer} } & \multicolumn{1}{c|}{\bf {\em Photo}}  & \multicolumn{1}{c|}{\bf {\em DBLP} } & \multicolumn{1}{c|}{\bf {\em PubMed}} & \multicolumn{1}{c|}{\bf {\em Cora}} & \multicolumn{1}{c}{\bf {\em ArXiv}}\\
	\hline
    $\alpha$ &$0.8$&$1.7$&$0.8$&$1.5$&$0.9$&$1.8$&$0.9/1.4$&$1.4/2.5$\\
    $T$ &$15$&$6/12$&$60/40$&$9$&$10$&$175$&$20/7$&$30$  \\
    $\tau$ &$7/50$&$7/50$&$7/100$&$7/50$&$7/50$&$7/50$&$7/100$ &$4/50$\\
    $\gamma$ &$1$&$0.9$&$0.9$&$1$&$1$&$1$&$1$&$1$ \\
    \hline
\end{tabular}%
}
\end{small}
\vspace{-1ex}
\end{table}

\begin{figure*}[t!]
\vspace{-3ex}
\centering
\resizebox{\textwidth}{!}{%
\subfloat[\small Ground truth]{
\includegraphics[width=6cm]{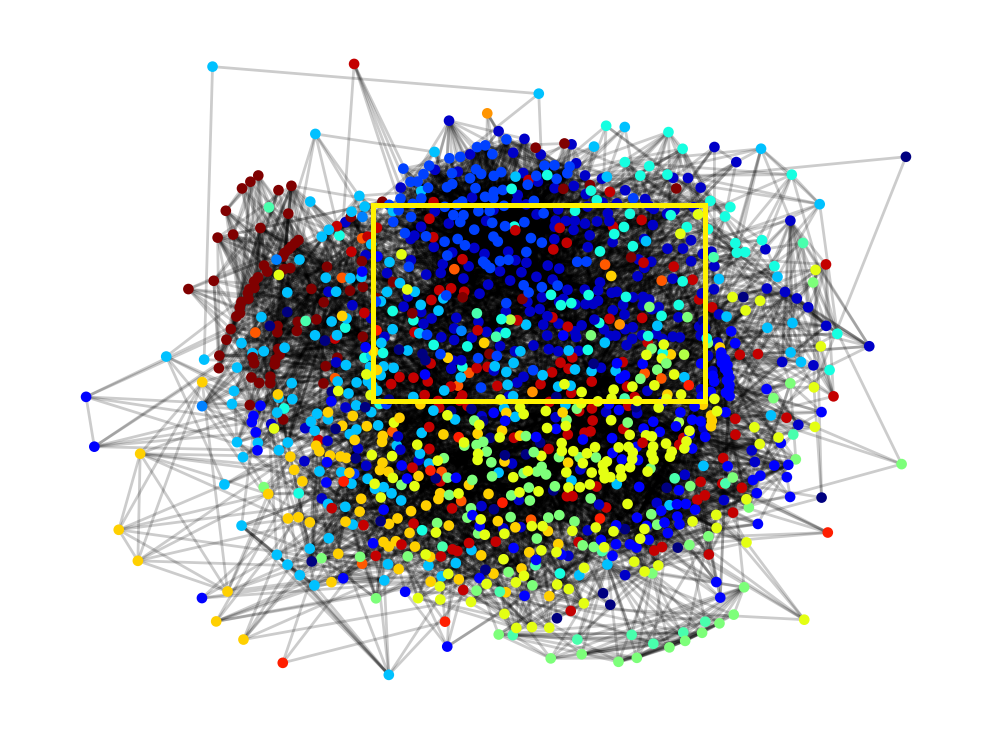}
\label{fig:wiki-groung}
}
\subfloat[\small \algo]{
\includegraphics[width=6cm]{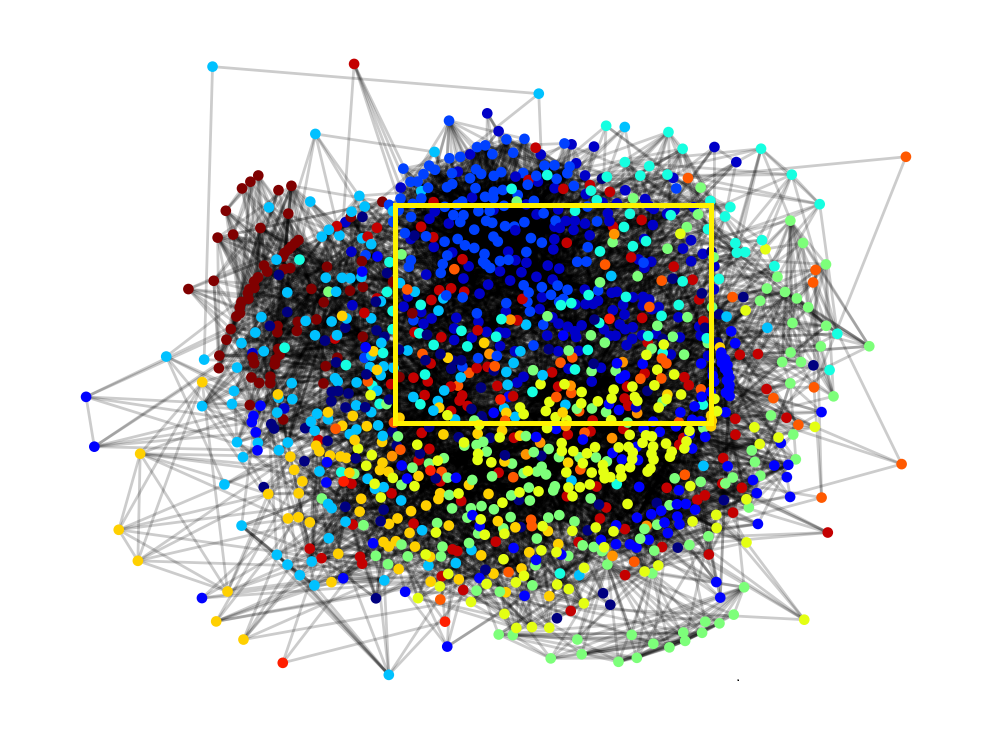}
\label{fig:wiki-scan}
}
\subfloat[\small \textsf{SAGSC}]{
\includegraphics[width=6cm]{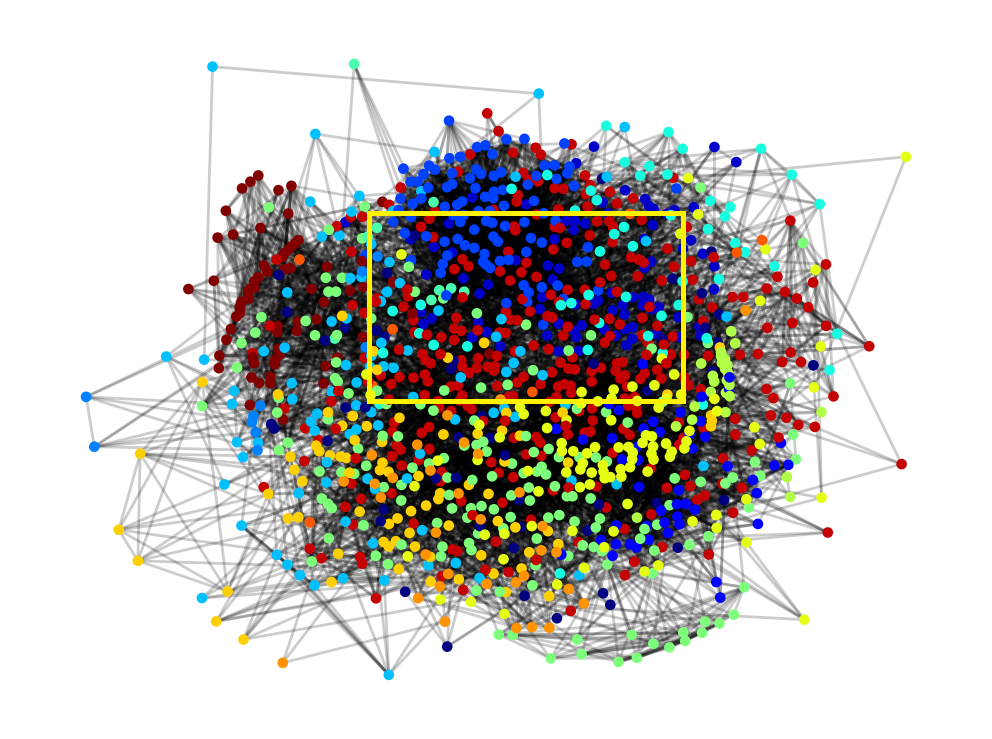}
\label{fig:wiki-sagsc}
}
}
\vspace{-4mm}
\caption{Visualizations on {\em Wiki}.}\label{fig:vir-wiki}
\vspace{-2ex}
\end{figure*}

\begin{figure*}[t!]
\vspace{-3ex}
\centering
\resizebox{\textwidth}{!}{%
\subfloat[\small Ground truth]{
\includegraphics[width=6cm]{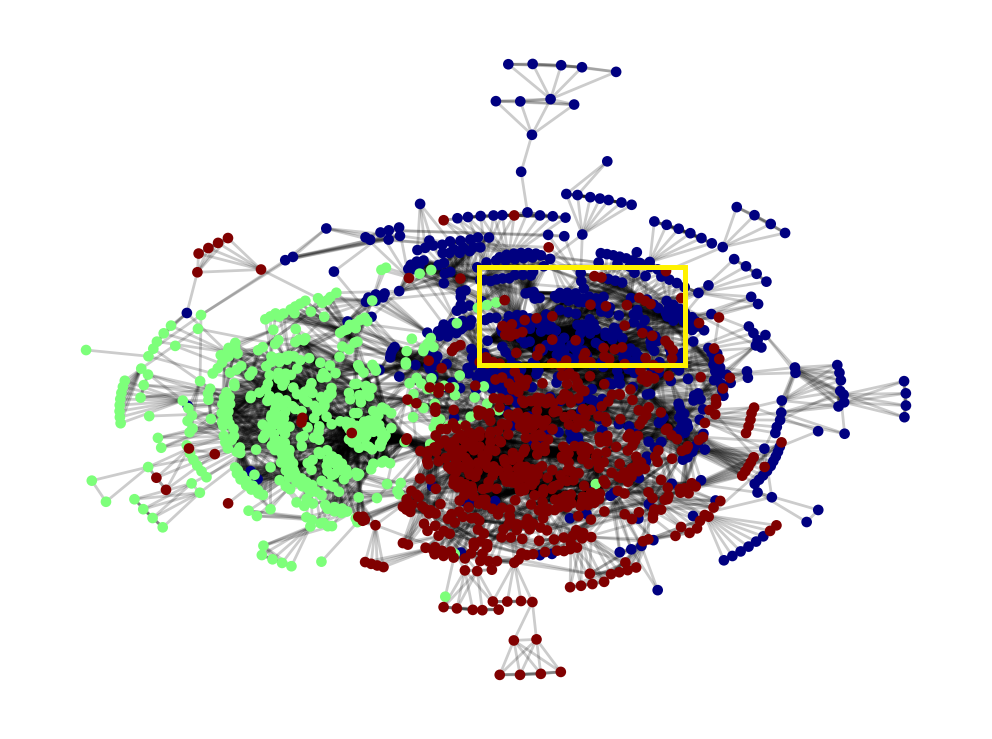}
\label{fig:acm-ground}
}
\subfloat[\small \algo]{
\includegraphics[width=6cm]{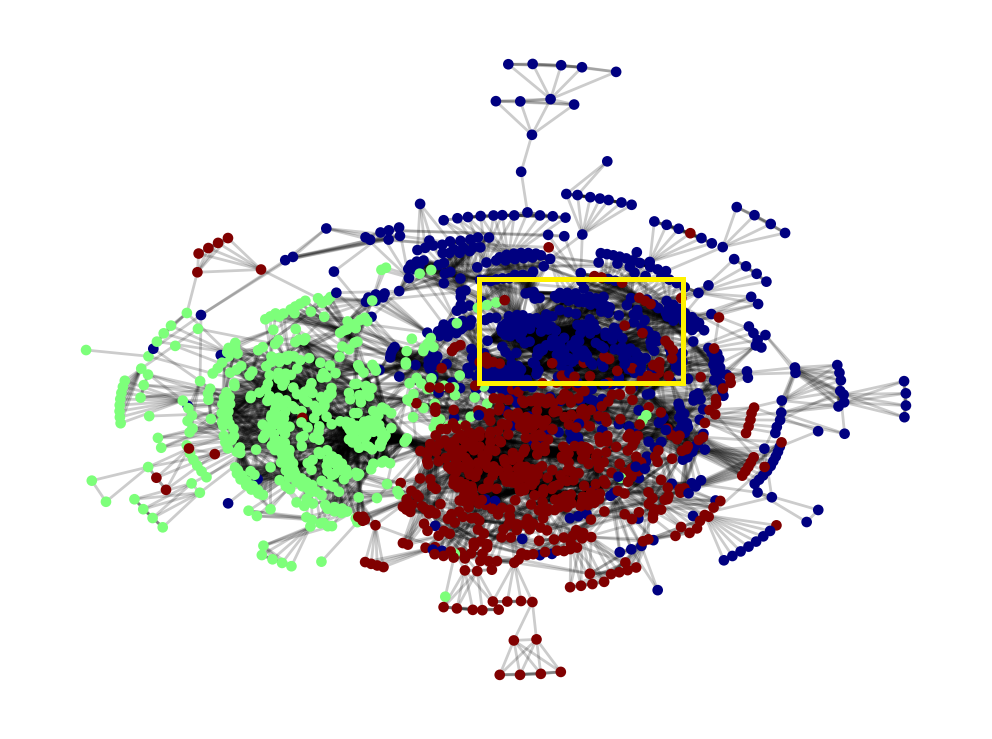}
\label{fig:acm-scan}
}
\subfloat[\small \textsf{SAGSC}]{
\includegraphics[width=6cm]{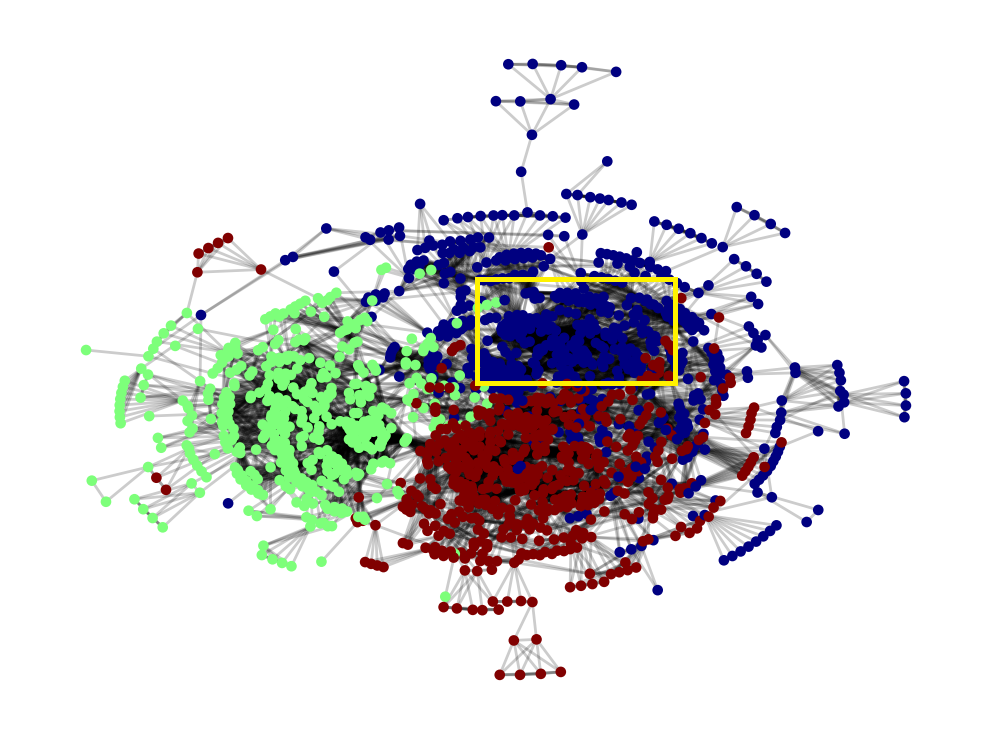}
\label{fig:acm-sagsc}
}
}
\vspace{-4mm}
\caption{Visualizations on {\em ACM}. }\label{fig:vir-acm}
\vspace{-2ex}
\end{figure*}

\begin{figure*}[t!]
\vspace{-3ex}
\centering
\resizebox{\textwidth}{!}{%
\subfloat[\small Ground truth]{
\includegraphics[width=6cm]{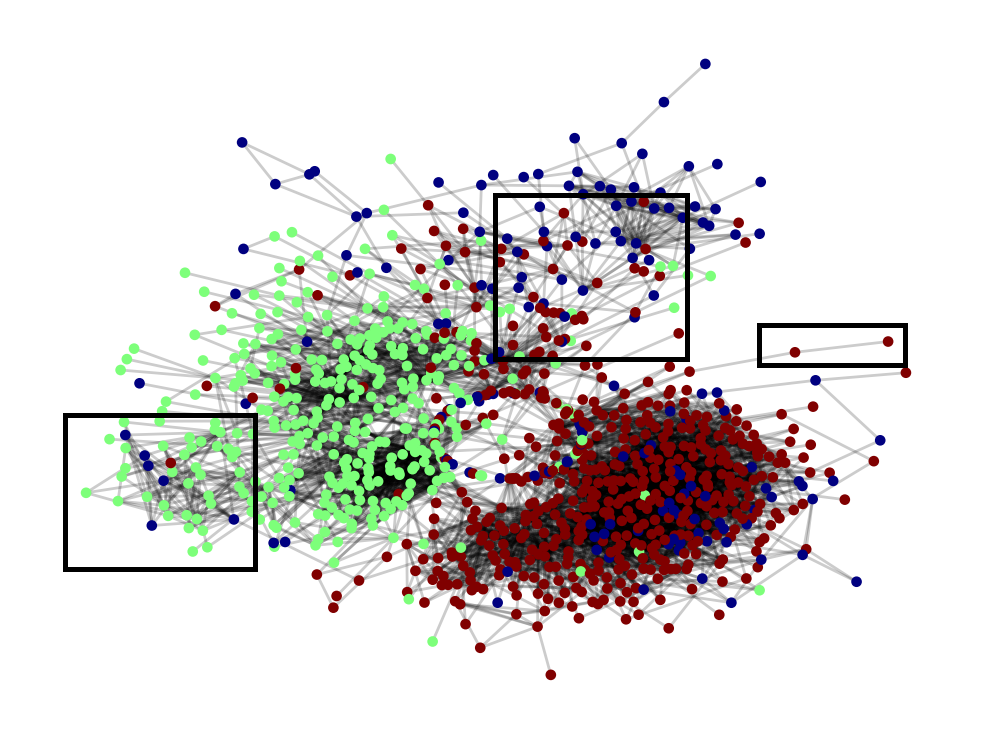}
\label{fig:pub-groung}
}
\subfloat[\small \algo]{
\includegraphics[width=6cm]{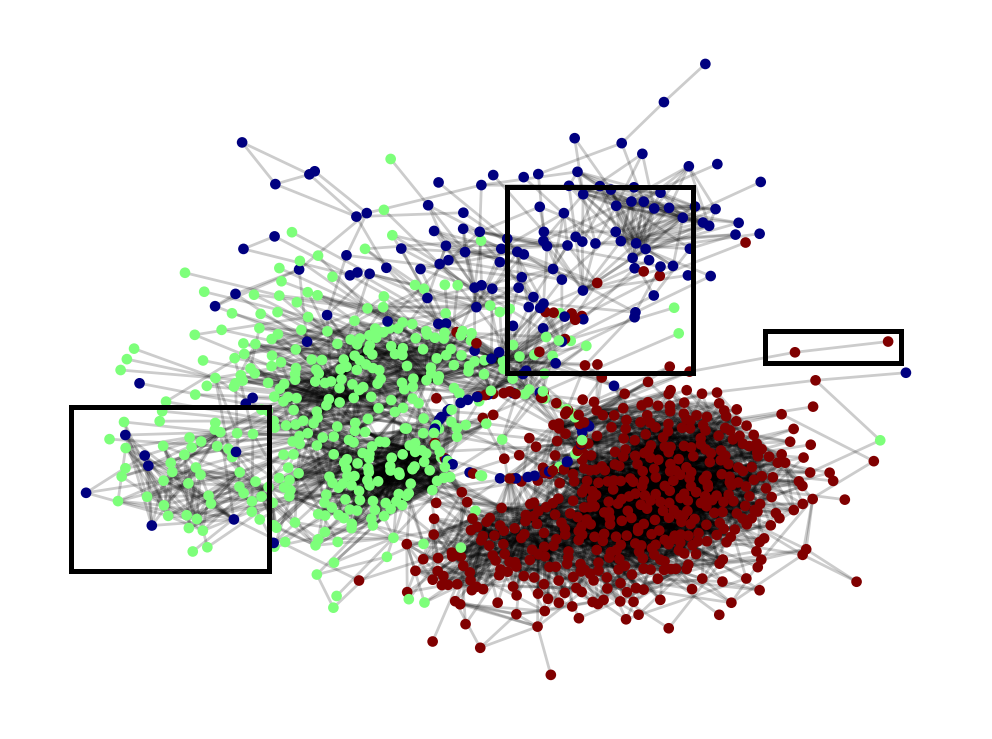}
\label{fig:pub-scan}
}
\subfloat[\small \textsf{SAGSC}]{
\includegraphics[width=6cm]{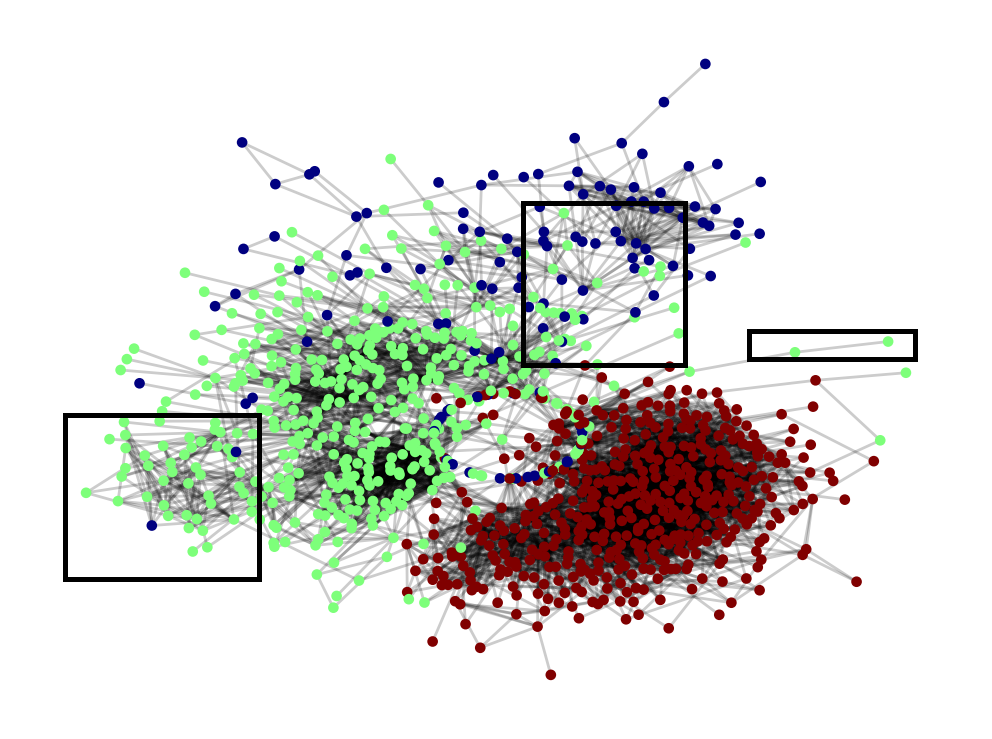}
\label{fig:pub-sagsc}
}
}
\vspace{-4mm}
\caption{Visualizations on {\em PubMed}.}\label{fig:vir-pubmed}
\vspace{-2ex}
\end{figure*}

\subsection{Clustering Visualization}
We visually compare the clustering results output by \algo and \textsf{SAGSC} against the ground truth on {\em Wiki}, {\em ACM}, and {\em PubMed} datasets. Specifically, for each of them, we run the well-known Fruchterman-Reingold force-directed algorithm to draw the layout of the graph, wherein vertices are colored as per their respective cluster labels (true ones or predicted ones).

Fig.~\ref{fig:vir-acm}, \ref{fig:vir-wiki}, \ref{fig:vir-pubmed} display the visualization results of the ground truth, \algo and \textsf{SAGSC} on {\em Wiki}, {\em ACM}, and {\em PubMed}, respectively. The major differences between the predicted results and the ground truth are highlighted in the rectangles. It can be observed that compared to \textsf{SAGSC}, the vertices in the highlighted areas are colored (i.e., assign cluster labels to vertices) in a way more similar to the ground truth by \algo, which is consistent with the fact that \algo enjoys higher clustering accuracy. For example, in Fig.~\ref{fig:vir-pubmed}, in the three rectangles, most of the vertices are mistakenly colored in green by \textsf{SAGSC}, whereas our \algo can color them in red and blue correctly. Accordingly, as reported in Table~\ref{tbl:node-clustering-large}, on {\em PubMed}, \algo outperforms \textsf{SAGSC} remarkably by a substantial margin of $4.2\%$, $3.6\%$, and $7.8\%$ for ACC, NMI, and ARI, respectively.

\section{Theoretical Proofs}\label{sec:proof}

\begin{proof}[\bf Proof of Lemma~\ref{lem:op}]
Suppose that $\boldsymbol{\Omega}^{\top}\boldsymbol{\Omega}= \IM$. Then,
\begin{align*}
\|\boldsymbol{\Omega}\MM-\NM\|^2_F &= \langle \boldsymbol{\Omega}\MM-\NM, \boldsymbol{\Omega}\MM-\NM\rangle_F \\
& = \|\boldsymbol{\Omega}\MM\|^2_F + \|\NM\|^2_F - 2 \langle\boldsymbol{\Omega}\MM,\NM \rangle_F \\
& = \|\MM\|^2_F + \|\NM\|^2_F - 2 \langle\boldsymbol{\Omega}\MM,\NM \rangle_F \\
& = \|\MM\|^2_F + \|\NM\|^2_F - 2 trace(\NM\MM^{\top}\boldsymbol{\Omega}^{\top}).
\end{align*}
Hence, the minimization of $\|\boldsymbol{\Omega}\MM-\NM\|^2_F$ is equivalent to maximizing $trace(\NM\MM^{\top}\boldsymbol{\Omega}^{\top})$. Let $\UM\boldsymbol{\Sigma}\VM^{\top}$ be the full SVD of $\NM\MM^{\top}$.
Let $\TM=\UM^{\top}\boldsymbol{\Omega}\VM$, which is an orthogonal matrix since $\TM^{\top}\TM = \IM$. This further implies that $-1\le \TM_{i,i}\le 1\ \forall{i}\in [1,n]$. 
Next, using the properties of matrix trace, we derive 
\begin{align*}
 trace(\NM\MM^{\top}\boldsymbol{\Omega}^{\top}) & = trace(\UM\boldsymbol{\Sigma}\VM^{\top}\boldsymbol{\Omega}^{\top}) =  trace(\UM^{\top}\boldsymbol{\Omega}\VM\boldsymbol{\Sigma}) \\
&  = trace(\TM\boldsymbol{\Sigma}) = \sum_{i=1}^{n}{\TM_{i,i}\cdot \boldsymbol{\Sigma}_{i,i}} \le \sum_{i=1}^{n}{\boldsymbol{\Sigma}_{i,i}},
\end{align*}
meaning that the trace $trace(\NM\MM^{\top}\boldsymbol{\Omega}^{\top})$ is maximized when $\TM=\IM$.
To achieve this, we need to find a rank-$k$ matrix $\boldsymbol{\Omega}$ minimizing 
$$\|\TM-\IM\|_F^2 = \| \UM^{\top}\boldsymbol{\Omega}\VM -\IM\|_F^2.$$
Since the columns of $\UM$ (resp. $\VM$) are singular vectors that are orthonormal, the problem can be rewritten as 
$$\min_{rank(\boldsymbol{\Omega})=k}\|\boldsymbol{\Omega}-\UM\VM^{\top}\|_F^2,$$ 
which is a classic {\em low-rank approximation} problem. 
By Eckart–Young theorem~\cite{horn2012matrix}, the solution can be derived through a $k$-truncated SVD over $\UM\VM^{\top}$, implying that $\boldsymbol{\Omega}=\UM^{(k)}{\VM^{(k)}}^{\top}$.
\end{proof}

\begin{proof}[\bf Proof of Lemma~\ref{lem:obj-rank}]
\eat{
Eckart–Young-Mirsky theorem~\cite{horn2012matrix} indicates that the optimal solution to Eq.~\eqref{eq:obj-rank} is $\SM=\UM^{\prime}\boldsymbol{\Sigma}^{\prime}{\VM^{\prime}}^{\top}$, where $\boldsymbol{\Sigma}^{\prime}$ stands for the diagonal matrix containing the top-$k$ singular values of $\UM\VM^{\top}$ and the columns in $\UM^{\prime}$ and ${\VM^{\prime}}$ are the corresponding left and right singular vectors, respectively. Since the columns in $\UM$ and $\VM$ are singular vectors, they are unitary matrices, whose product $\UM\VM^{\top}$ is also a unitary matrix. Recall that the singular values of a unitary matrix are all 1~\cite{horn2012matrix}, which implies that the left and right singular vectors of $\UM\VM^{\top}$ are $\UM$ and $\VM$, respectively.
Hence, we have $\UM^{(k)}=\UM^{\prime}$, $\VM^{(k)}={\VM^{\prime}}$, and $\SM=\UM^{(k)}{\VM^{(k)}}^{\top}$, where $\UM^{(k)}$ (resp. $\VM^{(k)}$) contains the top-$k$ left (resp. right) singular vectors of $\ZM\ZM^{\top}$.
}
\begin{theorem}[~\cite{horn2012matrix}]\label{lem:singular-eigs}
Let $\MM$ be an $n\times n$ real symmetric matrix. Let columns in $\boldsymbol{\Gamma}$, $\boldsymbol{\Upsilon}$, and $\boldsymbol{\Psi}$ be the left singular vectors, right singular vectors, and eigenvectors, respectively. Diagonal matrices $\boldsymbol{\Phi}$ and $\boldsymbol{\Lambda}$ consist of the singular values and eigenvalues of $\MM$ at their diagonal entries, respectively. Then, $\boldsymbol{\Gamma}_{\cdot,i}=\boldsymbol{\Psi}_{\cdot,i}$, $\boldsymbol{\Upsilon}_{\cdot,i}=\boldsymbol{\Psi}_{\cdot,i}\cdot sign(\boldsymbol{\Lambda}_{i,i})$, and $\boldsymbol{\Phi}_{i,i}=|\boldsymbol{\Lambda}_{i,i}|$ hold for $1\le i\le n$, where $sign(\cdot)$ stands for the sign function, i.e., $sign(x)=1$ if $x>0$, $sign(x)=0$ if $x=0$, and $sign(x)=-1$ if $x<0$.
\end{theorem}

Let $\UM^{(k)}$ (resp. $\VM^{(k)}$) contain the top-$k$ left (resp. right) singular vectors of $\ZM\ZM^{\top}$.
Next, we prove that $\UM^{(k)}=\VM^{(k)}$ and it is the top-$k$ left singular vectors of $\ZM$. 
Let the columns in $\UM^{\ast}$, $\VM^{\ast}$, and the diagonal entries in $\boldsymbol{\Sigma}^{\ast}$ be the left singular vectors, right singular vectors, and singular values of $\ZM$, respectively. Using the relation of SVD to eigendecomposition~\cite{horn2012matrix}, the columns of $\UM^{\ast}$ are eigenvectors of $\ZM\ZM^{\top}$ and the diagonal elements of $\boldsymbol{\Sigma}^{\ast}$ are the square roots of the eigenvalues of $\ZM\ZM^{\top}$. In other words, $\UM^{\ast}{\boldsymbol{\Sigma}^{\ast}}^2{\UM^{\ast}}^{\top}$ is the eigendecomposition of $\ZM\ZM^{\top}$.
Further, according to Theorem~\ref{lem:singular-eigs}, $\UM$ is equal to the eigenvectors of $\ZM\ZM^{\top}$, and thus, $\UM=\UM^{\ast}$. 

Note that singular values in $\boldsymbol{\Sigma}^{\ast}$ are non-negative. The top-$k$ singular vectors in $\UM^{\ast}$ of $\ZM$ are then exactly the $k$-largest eigenvectors of $\ZM\ZM^{\top}$ since its eigendecomposition is $\UM^{\ast}{\boldsymbol{\Sigma}^{\ast}}^2{\UM^{\ast}}^{\top}$. By Theorem~\ref{lem:singular-eigs} and the non-negativity of ${\boldsymbol{\Sigma}^{\ast}}^2$, the eigenvalues ${\boldsymbol{\Sigma}^{\ast}}^2$ of $\ZM\ZM^{\top}$ are the same as its singular values $\boldsymbol{\Sigma}$ and $\VM=\VM^{\ast}$. Accordingly, $\UM^{\ast}{\boldsymbol{\Sigma}^{\ast}}^2{\UM^{\ast}}^{\top}$ is the full SVD of $\ZM\ZM^{\top}$, namely, $\UM^{(k)}=\VM^{(k)}$ is the top-$k$ left singular vectors 
of $\ZM$, which proves the lemma.
\end{proof}

\begin{proof}[\bf Proof of Lemma~\ref{lem:Y-Uk}]
Recall that the columns of $\UM^{(k)}\in \mathbb{R}^{n\times k}$ are the top-$k$ left singular vectors of $\ZM$. Hence, the columns in $\UM^{(k)}$ are orthonormal.
Consider any column $\UM^{(k)}_{\cdot,i}$ of $\UM^{(k)}\ \forall{1\le i\le k}$. We have
\begin{equation*}
\UM^{(k)}{\UM^{(k)}}^{\top}\cdot \UM^{(k)}_{\cdot,i} = \UM^{(k)}\boldsymbol{e}_i = \UM^{(k)}_{\cdot,i},
\end{equation*}
where $\boldsymbol{e}_i\in \mathbb{R}^k$ stands for a column vector with value 1 at the $i$-th position and 0 elsewhere. The above equation implies that columns of $\UM^{(k)}$ are all eigenvectors of $\UM^{(k)}{\UM^{(k)}}^{\top}$ whose corresponding eigenvalues are $1$.

By the definition, $\UM^{(k)}{\UM^{(k)}}^{\top}$ is a rank-$k$ projection matrix onto the subspace spanned by the columns of $\UM^{(k)}$, whose eigenvalues are either 0 or 1.
Since its rank is $k$, i.e., the sum of all eigenvalues is $k$, $\UM^{(k)}{\UM^{(k)}}^{\top}$ has $k$ eigenvectors corresponding to eigenvalue 1 and other eigenvectors not in the span of \( U_k \) correspond to the eigenvalue 0. Consequently, we can conclude that the columns of $\UM^{(k)}$ are the $k$-largest eigenvectors of $\UM^{(k)}{\UM^{(k)}}^{\top}$ and the lemma is proved.
\end{proof}

\begin{proof}[\bf Proof of Theorem~\ref{lem:sscag}]
Let $\RM^{(0)}$ be the matrix $\RM$ generated at Line 5. Consider any iteration of Lines 7-9 in Algo.~\ref{alg:sscag}. Based on Eq.~\eqref{eq:NSR}, the output $\HM$ at Line 7 is $\sum_{t=0}^{T}{\frac{(1-\alpha)\alpha^{t}}{1-\alpha^{T+1}}\hat{\PM}^{t}}\hat{\XM}\RM=\ZM\RM$ and Line 8 returns 
$$\sum_{t=0}^{T}{\frac{(1-\alpha)\alpha^{t}}{1-\alpha^{\TM+1}}\hat{\PM^\top}^{t}}\HM=\sum_{t=0}^{T}{\frac{(1-\alpha)\alpha^{t}}{1-\alpha^{\TM+1}}\hat{\PM^\top}^{t}}\ZM\RM.$$
Accordingly, Line 9 yields 
$$\RM=\hat{\XM}^{\top}\HM = \hat{\XM}^{\top}\sum_{t=0}^{T}{\frac{(1-\alpha)\alpha^{t}}{1-\alpha^{\TM+1}}\hat{\PM^\top}^{t}}\ZM\RM=\ZM^{\top}\ZM\RM.$$
After repeating the above matrix multiplications for $\tau$ times, we have $\RM = (\ZM^{\top}\ZM)^{\tau}\RM^{(0)}$ at the end of $\tau$-th iteration. Line 10 furthers derives 
$$\HM = \sum_{t=0}^{T}{\frac{(1-\alpha)\alpha^{t}}{1-\alpha^{\TM+1}}\hat{\PM}^{t}}\cdot \hat{\XM}\RM = \ZM \cdot (\ZM^{\top}\ZM)^{\tau}\RM^{(0)} = (\ZM\ZM^{\top})^{\tau}\ZM\RM^{(0)}.$$

$\BM$ is calculated as at Line 12, which equals $\sum_{t=0}^{T}\frac{(1-\alpha)\alpha^{t}}{1-\alpha^{T+1}}{\hat{\PM^\top}^{t}}\QM$ and yields 
$$\textstyle \BM^{\top}\hat{\XM}=\QM^{\top}\sum_{t=0}^{T}\frac{(1-\alpha)\alpha^{t}}{1-\alpha^{T+1}}{\hat{\PM}^{t}}\hat{\XM}=\ZM\hat{\XM}$$.
Let $\boldsymbol{\Gamma}\boldsymbol{\Sigma}\VM^{\top}$ be the SVD of $\BM^{\top}\hat{\XM}.$
According to Theorem 1.2 in \cite{Halko2009FindingSW}, we have 
\begin{align*}
 \mathbb{E}\|\ZM-\YM^{\prime}\boldsymbol{\Sigma} \VM^\top\|\leq \left(1+4\sqrt{\frac{2\min\{n,d\}}{(k+o)/2-1}}\right)^{\frac{1}{2\tau+1}}\cdot\sigma_{(k+o)/2+1},
\end{align*}
where $\sigma_{(k+o)/2+1}$ signifies the $((k+o)/2+1)$-th largest singular value of $\ZM$. In sum, the columns in $\YM$ are the approximate top-$(k+o)$ left singular vectors of $\ZM$.
\end{proof}
\begin{proof}[\bf Proof of Lemma~\ref{lem:Y-trace}]
Using Ky Fan’s trace maximization principle~\cite{Fan1949OnAT}, the optimal solution to the following trace maximization problem $\max_{\boldsymbol{\Upsilon}\in \mathbb{R}^{n\times k}} trace(\boldsymbol{\Upsilon}^{\top}\ZM\ZM^{\top}\boldsymbol{\Upsilon})\ \text{subject to $\boldsymbol{\Upsilon}^{\top}\boldsymbol{\Upsilon}=\IM$}$ is the $k$-largest eigenvectors of $\ZM\ZM^{\top}$. Since $\YM$ is the top-$k$ left singular vectors of $\ZM$ and $\ZM\ZM^{\top}$ is a symmetric matrix, Theorem~\ref{lem:singular-eigs} states that $\YM$ is also the $k$-largest eigenvectors of $\ZM\ZM^{\top}$. The lemma is proved.
\end{proof}

\begin{proof}[\bf Proof of Lemma~\ref{lem:conductance}]
Since $\beta\cdot\WM$ is a stochastic matrix, then $\frac{\IM}{\beta}-\WM$ is the Laplacian of $\tilde{\G}$. For any row vector $\xvec \in \mathbb{R}^n$, it is easy to verify that 
$$\xvec^{\top}(\frac{\IM}{\beta}-\WM)\xvec = \frac{1}{2}\sum_{v_i,v_j\in \V}{\WM_{i,j}\cdot (\xvec_i-\xvec_j)^2}.$$
Let $\{\C_1,\C_2,\ldots,\C_k\}$ be the clusters corresponding to the VCA matrix $\CM$. By the definition of $\CM$ and its orthogonality property $\CM^{\top}\CM=\IM$, we deduce
\begin{small}
\begin{align*}
& trace(\CM^{\top}\WM\CM) = k - trace\left(\CM^{\top}\left(\frac{\IM}{\beta}-\WM\right)\CM\right) \\
 & = k-\frac{1}{2}\sum_{\ell=1}^{d}\sum_{v_i,v_j\in \V}{\WM_{i,j}\cdot (\CM_{i,\ell}-\CM_{j,\ell})^2} \\
& = k-\frac{1}{2}\sum_{\ell=1}^{d}\sum_{v_i\in \C_\ell,v_j\notin \C_\ell}{\frac{\WM_{i,j}}{|\C_\ell|}} = k-\frac{\phi(\C_1,\C_2,\ldots,\C_k)}{2}.
\end{align*}
\end{small}
The above equation indicates that the maximization of $trace(\CM^{\top}\WM\CM)$ is equivalent to the minimization of $\phi(\C_1,\C_2,\ldots,\C_k)$, which completes the proof.
\end{proof}

\begin{proof}[\bf Proof of Lemma~\ref{lem:mod-trace}]
First, let $\BM=\hat{\ZM}\hat{\ZM}^{\top} - \gamma \cdot \frac{\wvec\wvec^{\top}}{\wvec^{\top}\boldsymbol{1}}$. Consider any cluster $\C_\ell\in \{\C_1,\C_2,\ldots,\C_k\}$. By the definition of $\CM$, it is easy to verify that
\begin{align*}
\CM_{\cdot,\ell}^{\top}\BM\CM_{\cdot,\ell} = \sum_{v_i,v_j\in \C_\ell}{\BM_{i,j}}.
\end{align*}
For all the $k$ clusters, we have
\begin{align*}
trace(\CM^{\top}\BM\CM) = \sum_{\ell=1}^{k}\CM_{\cdot,\ell}^{\top}\BM\CM_{\cdot,\ell} = \sum_{\ell=1}^{k}\sum_{v_i,v_j\in \C_\ell}{\BM_{i,j}} = \wvec^{\top}\boldsymbol{1}\cdot Q,
\end{align*}
which leads to the lemma.
\end{proof}

\begin{proof}[\bf Proof of Theorem~\ref{lem:m-sscag}]
Since Eq.~\eqref{eq:update-H-Q} is equivalent to $\HM \gets (\hat{\ZM}\hat{\ZM}^{\top} - \gamma \cdot \frac{\wvec\wvec^{\top}}{\wvec^{\top}\boldsymbol{1}})\cdot \QM$, by Theorem 5.1 in \cite{saad2011numerical}, the columns of $\QM$ are the $k$-largest eigenvectors $\YM$ of $\hat{\ZM}\hat{\ZM}^{\top} - \gamma \cdot \frac{\wvec\wvec^{\top}}{\wvec^{\top}\boldsymbol{1}}$ when $\QM$ converges. 

Next, suppose that matrix $\tilde{\ZM}$ satisfies $\textstyle \tilde{\ZM}\tilde{\ZM}^{\top} = \hat{\ZM}\hat{\ZM}^{\top} - \gamma \cdot \frac{\wvec\wvec^{\top}}{\wvec^{\top}\boldsymbol{1}}$. According to Lemma~\ref{lem:op} and Lemma~\ref{lem:obj-rank}, the solution $\SM$ to the problem in Eq.~\eqref{eq:obj-M} is $\SM=\UM^{(k)}{\UM^{(k)}}^{\top}$, where $\UM^{(k)}$ consists of the top-$k$ left singular vectors of $\tilde{\ZM}$. Similar to the proof of Lemma~\ref{lem:obj-rank}, using Theorem~\ref{lem:singular-eigs} derives that the $k$-largest eigenvectors $\YM$ of $\tilde{\ZM}$ are the top-$k$ left singular vectors of $\tilde{\ZM}$, namely $\SM=\YM\YM^{\top}=\QM\QM^{\top}$, which completes the proof.
\end{proof}

\begin{proof}[\bf Proof of Lemma~\ref{lem:zz-range}]
Let $\textstyle \PiM = \sum_{t=0}^{T}{\frac{(1-\alpha)\alpha^t}{1-\alpha^{T+1}}\hat{\PM}^{t}}$. We prove that $\PiM$ is a right stochastic matrix.
According to the definition of $\hat{\PM}$ in Section~\ref{sec:SCAG}, $\forall{v_i\in \V}$, $\sum_{v_j\in \V}\hat{\PM}_{i,j}=1$ holds, meaning that $\hat{\PM}$ is right stochastic, and thus, $\hat{\PM}^t\ \forall{t\ge 0}$ is right stochastic. For any vertex $v_i\in \V$,
\begin{small}
\begin{align*}
\sum_{v_j\in \V}\PiM_{i,j} & = \sum_{v_j\in\V}\sum_{t=0}^{T}{\frac{(1-\alpha)\alpha^t}{1-\alpha^{T+1}}\hat{\PM}^{t}_{i,j}} = \sum_{t=0}^{T}{\frac{(1-\alpha)\alpha^t}{1-\alpha^{T+1}}\sum_{v_j\in\V}\hat{\PM}^{t}_{i,j}} \\
& = \sum_{t=0}^{T}{\frac{(1-\alpha)\alpha^t}{1-\alpha^{T+1}}} = 1,
\end{align*}
\end{small}
indicating that $\PiM$ is right stochastic.

Next, we denote $\hat{\XM}\hat{\XM}^{\top}$ by $\BM$.
We can represent $\ZM\ZM^{\top}$ as $\PiM \BM\PiM^{\top}$, and hence, for $v_i\in \V$, 
\begin{small}
\begin{align*}
 \sum_{v_j\in \V}\ZM_i{\ZM_j}^{\top} &  = \sum_{v_j\in \V}\sum_{v_h\in \V}\sum_{v_\ell\in \V}\PiM_{i,\ell}\cdot \BM_{\ell,h}\cdot \PiM_{j,h} \\
& = \sum_{v_j\in \V}\sum_{v_\ell\in \V}\PiM_{i,\ell} \sum_{v_h\in \V} \BM_{\ell,h}\cdot \PiM_{j,h}\\
& = \sum_{v_\ell\in \V}\PiM_{i,\ell} \sum_{v_h\in \V} \BM_{\ell,h}\sum_{v_j\in \V}\PiM_{j,h}.
\end{align*}
\end{small}
Then, by the definitions of $\beta_\ell$ and $\pi_\ell$, 
\begin{small}
\begin{equation*}
\begin{aligned}
\sum_{v_\ell\in \V}\PiM_{i,\ell} \sum_{v_h\in \V} \BM_{\ell,h}\sum_{v_j\in \V}\PiM_{j,h} & \le \max_{v_\ell\in \V}{\beta_\ell\cdot \pi_\ell} \cdot \sum_{v_\ell\in \V}\PiM_{i,\ell}\\
\sum_{v_\ell\in \V}\PiM_{i,\ell} \sum_{v_h\in \V} \BM_{\ell,h}\sum_{v_j\in \V}\PiM_{j,h} & \ge \min_{v_\ell\in \V}{\beta_\ell\cdot \pi_\ell} \cdot \sum_{v_\ell\in \V}\PiM_{i,\ell}.
\end{aligned}
\end{equation*}
\end{small}
Since $\PiM$ is a right stochastic matrix, we can get
\begin{equation*}
\min_{v_\ell\in \V}{\beta_\ell\cdot \pi_\ell} \le \sum_{v_j\in \V}\ZM_i{\ZM_j}^{\top} \le \max_{v_\ell\in \V}{\beta_\ell\cdot \pi_\ell},
\end{equation*}
which finishes the proof.
\eat{
Let $b_{\min}=\min_{v_\ell,v_h\in \V}\BM_{\ell,h}$ and $b_{\max}=\min_{v_\ell,v_h\in \V}\BM_{\ell,h}$. We have
\begin{small}
\begin{align*}
\sum_{v_j\in \V}\sum_{v_\ell\in \V}\PiM_{i,\ell} \sum_{v_h\in \V} \BM_{\ell,h}\cdot \PiM_{j,h} & \le \sum_{v_j\in \V}b_{\max}\sum_{v_\ell\in \V}\PiM_{i,\ell} \sum_{v_h\in \V} \PiM_{j,h} \\
& = \sum_{v_j\in \V}b_{\max}\sum_{v_\ell\in \V}\PiM_{i,\ell} = n\cdot b_{\max}.
\end{align*}
\end{small}
and 
\begin{small}
\begin{align*}
\sum_{v_j\in \V}\sum_{v_\ell\in \V}\PiM_{i,\ell} \sum_{v_h\in \V} \BM_{\ell,h}\cdot \PiM_{j,h} & \ge \sum_{v_j\in \V}b_{\min}\sum_{v_\ell\in \V}\PiM_{i,\ell} \sum_{v_h\in \V} \PiM_{j,h} \\
& = \sum_{v_j\in \V}b_{\min}\sum_{v_\ell\in \V}\PiM_{i,\ell} = n\cdot b_{\min},
\end{align*}
\end{small}
}
\end{proof}

\begin{proof}[\bf Proof of Lemma~\ref{lem:XX-range}]
Let $\FM$ be $\XM\XM^{\top}$ and hence $\FM_{i,j}=\XM_i\cdot\XM_j^{\top}\ \forall{v_i,v_j\in \V}$. By the definition of $\hat{\XM}$ in Section~\ref{sec:SCAG},
\begin{small}
\begin{align*}
\hat{\XM}_i\cdot\hat{\XM}_j^{\top} = \frac{\FM_{i,j}}{\sqrt{\sum_{v_\ell\in \V}{\FM_{i,\ell}}}\sqrt{\sum_{v_\ell\in \V}{\FM_{j,\ell}}}}.
\end{align*}
\end{small}
Next, we can derive 
\begin{small}
\begin{align*}
\frac{1}{n^2}\sum_{v_i,v_j\in \V}{\hat{\XM}_i\cdot\hat{\XM}_j^{\top}} &= \frac{1}{n^2}\sum_{v_i,v_j\in \V}{\frac{\FM_{i,j}}{\sqrt{\sum_{v_\ell\in \V}{\FM_{i,\ell}}}\sqrt{\sum_{v_\ell\in \V}{\FM_{j,\ell}}}}}\\
& = \frac{1}{n^2}\sum_{v_i,v_j\in \V}{\sqrt{\frac{\FM_{i,j}}{\sum_{v_\ell\in \V}{\FM_{i,\ell}}}}\cdot \sqrt{\frac{\FM_{i,j}}{\sum_{v_\ell\in \V}{\FM_{j,\ell}}}}} 
\end{align*}
\end{small}
Using Cauchy–Schwarz inequality, we have
\begin{small}
\begin{align*}
& \frac{1}{n^2}\sum_{v_i,v_j\in \V}{\sqrt{\frac{\FM_{i,j}}{\sum_{v_\ell\in \V}{\FM_{i,\ell}}}}\cdot \sqrt{\frac{\FM_{i,j}}{\sum_{v_\ell\in \V}{\FM_{j,\ell}}}}}  \\
&\le \frac{1}{n^2}\sqrt{ \sum_{v_i,v_j\in \V}{\frac{\FM_{i,j}}{\sum_{v_\ell\in \V}{\FM_{i,\ell}}}}\cdot \sum_{v_i,v_j\in \V}{\frac{\FM_{i,j}}{\sum_{v_\ell\in \V}{\FM_{j,\ell}}}}}  \\
& = \frac{1}{n^2}\sqrt{ \sum_{v_i\in \V}{\frac{\sum_{v_j\in \V}\FM_{i,j}}{\sum_{v_\ell\in \V}{\FM_{i,\ell}}}}\cdot \sum_{v_j\in \V}{\frac{\sum_{v_i\in \V}\FM_{i,j}}{\sum_{v_\ell\in \V}{\FM_{j,\ell}}}}} = \frac{1}{n^2}\sqrt{n\cdot n} = \frac{1}{n}.
\end{align*}
\end{small}
The lemma is proved.
\end{proof}

\begin{proof}[\bf Proof of Lemma~\ref{lem:pisum}]
First, we denote 
$\sum_{t=0}^{T}{\frac{(1-\alpha)\alpha^t}{1-\alpha^{T+1}}\hat{\PM}^{t}}$ by $\PiM$. According to the definition of $\hat{\PM}$ in Section~\ref{sec:SCAG}, we can deduce that $\hat{\PM}=\hat{\DM}^{-1}\NAM$, where $\hat{\DM}$ is a diagonal matrix wherein each entry $\hat{\DM}_{h,h}$ equals $\hat{d}(v_h)$. Accordingly, $\PiM\hat{\DM}^{-1}$ is a symmetric matrix. This leads to $\textstyle \frac{\PiM_{h,\ell}}{\hat{d}(v_\ell)} = \frac{\PiM_{\ell,h}}{\hat{d}(v_h)}$.
Recall that $\PiM$ is a right stochastic matrix (see the proof of Lemma~\ref{lem:zz-range}), which implies $\textstyle \sum_{v_h\in \V}\PiM_{h,\ell} \cdot \frac{\hat{d}(v_h)}{\hat{d}(v_\ell)} = 1$.
Consequently,
\begin{small}
\begin{equation*}
\frac{\hat{d}(v_\ell)}{\max_{v_h\in \N_{v_\ell} \cap \{v_\ell\} }{\hat{d}(v_h)}} \le \sum_{v_h\in \V}\PiM_{h,\ell} \le \frac{\hat{d}(v_\ell)}{\min_{v_h\in \N_{v_\ell} \cap \{v_\ell\} }{\hat{d}(v_h)}}.
\end{equation*}
\end{small}
The lemma naturally follows.
\end{proof}

\end{document}